\documentclass[12pt,aps,prd,floats,preprint,preprintnumbers,nofootinbib]{revtex4}

\usepackage[dvips]{graphicx}
\usepackage{multirow}

\input{epsf.sty}
\def\fo{\hbox{{1}\kern-.25em\hbox{l}}}

\def\slashchar#1{\setbox0=\hbox{$#1$}           % set a box for #1
   \dimen0=\wd0                                 % and get its size 
   \setbox1=\hbox{/} \dimen1=\wd1               % get size of /
   \ifdim\dimen0>\dimen1                        % #1 is bigger 
      \rlap{\hbox to \dimen0{\hfil/\hfil}}      % so center / in box 
      #1                                        % and print #1
   \else                                        % / is bigger 
      \rlap{\hbox to \dimen1{\hfil$#1$\hfil}}   % so center #1
      /                                         % and print /
   \fi}                                         %

\def\beq{\begin{equation}}
\def\eeq{\end{equation}}
\def\mEt{\mbox{${\hbox{$E$\kern-0.6em\lower-.1ex\hbox{/}}}_T$}\, } %missing ET
\def\mpt{\mbox{${\hbox{$p$\kern-0.6em\lower-.1ex\hbox{/}}}_T$}\, } %missing pT

\def\snu{{\tilde\nu}}
\def\snuL{{\tilde\nu_L}}

\def\sNR{{\tilde{N}_R}}

\def\stR{{\tilde{t}_R}}
\def\sbR{{\tilde{b}_R}}
\def\sH{{\tilde H}}

\def\schi{{\tilde\chi}}
\def\sGl{{\tilde g}}

\newcommand{\newc}{\newcommand}

\newc{\mstop}{m_{\tilde{t}}}

\newc{\mtop}{m_t}
\newc{\mbot}{m_b}
\newc{\mz}{m_Z}
\newc{\mw}{M_W}

\newc{\sgn}{\mbox{sgn}}
\newc{\tbeta}{\tan\beta}

\newc{\Mlsp}{M_{\rm LSP}}

\def\beq{\begin{equation}}
\def\eeq{\end{equation}}
\def\bea{\begin{eqnarray}}
\def\eea{\end{eqnarray}}

\newcommand {\ksn} [1] {k^\nu_{#1}}
\newcommand {\ksnc} [1] {k^{\nu*}_{#1}}
\newcommand {\kt} [1] {k^t_{#1}}
\newcommand {\ktc} [1] {k^{t*}_{#1}}

\newcommand {\lsn} [1] {l^\nu_{#1}}
\newcommand {\lsnc} [1] {l^{\nu*}_{#1}}
\newcommand {\lt} [1] {l^t_{#1}}
\newcommand {\ltc} [1] {l^{t*}_{#1}}

\newcommand {\pslash} [1] { {p \hspace{-1.8mm} /}_{#1}}
\newcommand {\chip} [1] {\tilde{\chi}^+_{#1} }

\begin{document}

\begin{titlepage}

\begin{flushright}
%hep-ph/06mmnnn
IFIC/06-28 \\
NUHEP-TH/06-05 \\
%Version 4.0
\end{flushright}

\title{
Stop Decay into Right-handed Sneutrino LSP at Hadron Colliders\\
}

\author{Andr\'e de Gouv\^ea}
\email{degouvea@northwestern.edu}
\affiliation{Northwestern University, Department of Physics \& Astronomy, 2145 Sheridan Road, Evanston, IL~60208, USA.}

\author{Shrihari Gopalakrishna}
\email{shri@northwestern.edu}
\affiliation{Northwestern University, Department of Physics \& Astronomy, 2145 Sheridan Road, Evanston, IL~60208, USA.}

\author{Werner Porod}
\email{porod@ific.uv.es}
\affiliation{IFIC - Instituto de Fisica Corpuscular, CSIC, E-46071 Valencia, Spain.}

\date{\today}

\begin{abstract}

Right-handed neutrinos offer us the possibility of accommodating neutrino masses.
In a supersymmetric model, this implies the existence of right-handed sneutrinos.
Right-handed sneutrinos are expected to be as light as other supersymmetric particles  
if the neutrinos are 
Dirac fermions or if the lepton-number breaking scale is at (or below) the supersymmetry 
(SUSY) breaking scale, 
assumed to be around the electroweak scale. Depending on the mechanism of SUSY breaking,
 the lightest right-handed sneutrino may be the  lightest supersymmetric particle (LSP).
We consider the unique hadron collider signatures of 
a weak scale right-handed sneutrino LSP, assuming $R$-parity conservation.
For concreteness, we concentrate on stop pair-production and decay at the 
Tevatron and the 
Large Hadron Collider, and briefly comment on the production and decay of 
other supersymmetric particles.

\end{abstract}

\maketitle
\thispagestyle{empty}

\end{titlepage}

%%%%%%%%%%%%%%%%%%%%%%%%%%%%%%%%%%%%%%%%%%%%%%%%%%%%%%%%%%%%%%%%%%%%%
\section{Introduction}

A renormalizable extension of the standard model (SM)
that incorporates neutrino masses 
can be obtained by introducing (at least two) SM gauge singlet fermions -- right-handed neutrinos
$N_R$. Neutrino masses arise due to the fact that right-handed neutrinos can couple to the 
left-handed ones ($\nu_L$) through Yukawa couplings. After electroweak symmetry breaking,
the SM neutrinos are endowed with either Majorana or Dirac masses.
A supersymmetric version of this scenario implies the existence of
new, complex, SM gauge singlet scalar fields -- the
right-handed\footnote{``Sfermions'' are, of course, scalar particles, and have no sense of handedness. Throughout, however, we refer to left- and right-handed sfermions (scalar neutrinos, scalar tops, etc), as is commonly done in the literature, in order to indicate the super-partner of the various left- and right-handed chiral fermion fields (the neutrino, the top quark, etc).} sneutrinos $\sNR$.

The right-handed neutrino superfield $\hat N$ is a SM gauge singlet 
and is allowed to have a supersymmetry (SUSY) preserving Majorana mass $M_N$. 
In the usual high-scale  seesaw mechanism~\cite{Minkowski:1977sc},
$M_N$ -- the scale of lepton number breaking -- is, in general, not 
related to the SUSY breaking scale. It is often considered to be around $10^{14}~$GeV
so that, if the  neutrino Yukawa couplings are order one, active neutrino masses are
around $0.1~$eV, the scale inferred from neutrino oscillation experiments~\cite{osc_review}.
This scenario, while elegant, is very hard to verify experimentally\footnote{
One possibility is to look for right-handed neutrino traces in the RGE evolution of the Soft SUSY 
breaking parameters, which could be revealed using precision measurements at a next-generation 
linear collider~\cite{Blair:2002pg}.} and is motivated by the 
fact that neutrino Yukawa couplings are ``naturally'' expected to be of order one. If the seesaw 
energy scale is indeed very high, right-handed sneutrino masses are expected to be of order
$M_N$ and, hence, these are decoupled from low-energy phenomena.
SUSY effects in a theory with a high seesaw scale have been considered in 
\cite{SUSYseesaw}. 
It is also possible that neutrino masses arize from couplings to an SU(2) triplet Higgs 
boson~\cite{typeIIrefs}. The supersymmetric version of such a theory doesn't have to
include a right-handed neutrino superfield and thus its signatures is outside the purview of 
this work.
 
It has recently been emphasized that any value of $M_N$ is technically natural~\cite{deGouvea:2005er}
(including $M_N\equiv 0$, in which case the neutrinos are Dirac fermions), and that it is important
to explore the phenomenological consequences of all values of $M_N$. One intriguing
possibility is to consider, for example,  that the unknown physics of SUSY
breaking and lepton number breaking are intimately connected,
resulting in a common mass scale. 
If this is the case (or if the SUSY breaking scale were larger than $M_N$), right-handed sneutrino masses are expected to be  of
the same order as the masses of all other super-partners. Finally, 
the radiative stability of the weak
scale indicates that the SUSY breaking scale ought to be around the weak scale, so that super-partners
are expected to make their presence felt at TeV scale collider experiments. 

Depending on the physics of SUSY breaking, 
right-handed sneutrino masses may end up 
 below all other ``active'' super-partner masses, so that the lightest
sneutrino is the lightest supersymmetric partner (LSP).
In general, left- and right-handed sneutrinos mix. However, given that the seesaw
energy scale is assumed to be 
very low, the neutrino Yukawa coupling is, phenomenologically, required to be very small. Hence, if one assumes that the supersymmetric breaking ``A-terms'' 
are proportional to the respective Yukawa couplings, left-right sneutrino mixing is expected to be
very small (see, for example, \cite{Gopalakrishna:2006kr} for details, some aspects of which
are given in App.~\ref{Theory.SEC} for convenience) and the LSP will turn out to be 
composed of a mostly right-handed sneutrino, $\sNR$. 

In a recent work~\cite{Gopalakrishna:2006kr}, we considered such a $\sNR$ LSP when
the Majorana and SUSY breaking mass scales are around the electroweak
scale.  In this case, the neutrino Yukawa coupling $Y_N \sim 10^{-6}$
in order for the light neutrino masses to be to be of the order of
$0.1~$eV.  We analyzed the sneutrino sector, pointed out in which
cases the left-right sneutrino mixing angle can be tiny resulting in
an almost pure $\sNR$ LSP, and argued that the $\sNR$ is an interesting
non-thermal dark matter candidate\footnote{In case of additional non-standard 
interactions, it can also be a thermal dark matter candidate~\cite{Garbrecht:2006az}.} 
(see also \cite{Asaka:2005cn}). 
In this paper we explore the hadron
collider implications of a predominantly $\sNR$ LSP.  Collider
signatures of a mixed $\snuL-\sNR$ LSP have been explored previously in
other studies~\cite{SnuLRefs}.

Since a weak scale $\sNR$ LSP interacts only through the tiny Yukawa
coupling $Y_N$, many unique collider signatures are expected to arise.
 We will show that the most
striking among them is the possibility of the next-to-lightest
supersymmetric particle (NLSP) being long-lived enough to leave a
displaced vertex in the detector. A similar situation exists in
theories of gauge mediation where the gravitino is the
LSP~\cite{Chou:1999zb,GMSBRef}, with usually the scalar tau or the
lightest neutralino as NLSPs. 
A displaced vertex signature in the context of ``hidden valley'' models is 
discussed in \cite{Strassler:2006im}.
However, the scenario discussed here 
is distinguished by (a) the potentially long-lived states can be strongly interacting, (b) the
LSP carries lepton number, which implies associated leptons in the
final state, and (c) the LSP interacts only through the Yukawa
coupling $Y_N$, which generically implies non-universal rates for $e$,
$\mu$, and $\tau$ type leptons.

In order to illustrate the unique collider signatures of a $\sNR$ LSP, 
we consider in detail the case where the right-handed stop ($\stR$) is the NLSP\footnote{For
simplicity we will take the light $\tilde t$ to be predominantly $\stR$. While stop mixing is,
in general, not negligible, we find that its inclusion does not change, qualitatively, any of our results.}, 
and analyze its pair-production and decay.
While serving to illustrate many of the unique features of a $\sNR$ LSP, a light stop
is natural in many scenarios~\cite{LtStRef} and is favored in successful  
theories of electroweak baryogenesis~\cite{EWBaryoRefs}.
Stops, being strongly interacting particles, are also expected to be produced at significant rates at the 
Tevatron and the LHC.
Stop production and decay have also been analyzed in other 
contexts~\cite{Chou:1999zb,StopPDRef,Allanach:2006fy}. 

The outline of the paper is as follows:
In Sec.~\ref{stRdec.SEC}, we compute the decay rate of the $\stR$ into the 3-body
final state $b\ell^+ \sNR$. We incorporate this decay matrix element into the
Monte Carlo program Pythia (version 6.327)~\cite{PythiaRef}
 and study the Tevatron and LHC signatures 
in Sec.~\ref{stRProDec.SEC}. In Sec.~\ref{OthSusy.SEC}, we briefly discuss the dominant signatures
of other SUSY NLSP candidates, such as gluinos, sbottoms, gauginos, and sleptons, and
comment on 
``co-LSP'' right-handed sneutrinos. We offer our conclusions in Sec.~\ref{CONCL.SEC}.
The details of the model with which  we work were spelled out in 
detail in~\cite{Gopalakrishna:2006kr}, and are summarized in App.~\ref{Theory.SEC}.
The exact expression for the stop decay matrix element is given in
App.~\ref{AppDecay.SEC}.

%%%%%%%%%%%%%%%%%%%%%%%%%%%%%%%%%%%%%%%%%%%%%%%%%%%%%%%%%%%%%%%%%%%%%%%%%%%%%%%%%%
\section{$\stR$ decay to $\sNR$}
\label{stRdec.SEC}

In the scenario of interest, assuming $R$-parity conservation, 
all supersymmetric particles eventually cascade-decay to the
stable $\sNR$ LSP. Since
the $\sNR$ couples only via the tiny $Y_N$, the NLSP can be
potentially long-lived. To illustrate this aspect, in this section we
consider in detail the decay of the right-handed stop ($\stR$).
This will be particularly relevant when the $\stR$ is not too heavy, in which case
its production cross-section can be large enough to be observable at the
Tevatron or the LHC. A brief discussion of the decays of other NLSP candidates
is given in Sec.~\ref{OthSusy.SEC}.
We omit, for simplicity, stop mixing. We find that the  inclusion 
of left-right scalar top mixing does not change
significantly any of our results.

%%%%%%%%%%%%%%%%%%%%%%%%%%%%%%%%%%%%%%%%%%
\begin{figure}
\begin{center}
\scalebox{1.2}{\includegraphics{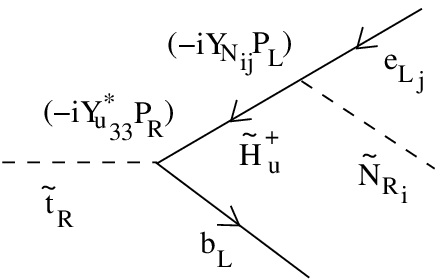}}
\caption{
$\stR\to b\ell\sNR$ decay mode. The arrows indicate fermion number flow.
\label{sTdecay.FIG}}
\end{center}
\end{figure}
Fig.~\ref{sTdecay.FIG} shows the dominant contribution to
the decay of a predominantly right-handed stop, which is assumed to be the NLSP, into
a pure $\sNR$ LSP.
The related mode $\tilde t \rightarrow t \nu \sNR$ can also be relevant
if kinematically allowed, although phase-space suppressed with respect to
$\tilde{t}\to b\ell\sNR$ due to the top quark in
the final state.

The complete expression for the lightest stop decay 
matrix element, including the mixings of all relevant SUSY particles,  is presented in
App.~\ref{AppDecay.SEC}.
For a purely right-handed stop and  sneutrino, 
in the limit of pure higgsino $\tilde H$ exchange\footnote{We find that, for the 
intentions of this paper, it is safe to ignore higgsino--gaugino mixing.}, 
the formula for the stop decay width has a simple form. 
In the $\stR$ rest frame, defining $\theta_{b\ell}$ as the angle between the 
bottom quark and the charged lepton, the matrix element squared is given by 
\beq
|T_{fi}|^2 \sim \frac{4 |Y_t|^2 |Y_N|^2 M_\stR^2 E_b E_\ell}{\left((p_\stR - k_b)^2 - M_{\tilde H}^2 \right)^2}
\frac{\left(1+\cos{\theta_{b\ell}}\right)}{2} \ .
\label{TfistR.EQ}
\eeq
Here, $|Y_t|$ ($|Y_N|$) is the top (neutrino) Yukawa coupling, $E_b$ ($E_{\ell}$) is the $b$-quark (charged lepton) energy, while $k_b$ is the $b$-quark four-momentum. $M_{\stR}$ and $p_\stR$ are, respectively, the $\stR$ mass and four-momentum, while $M_{\tilde{H}}$ is the Higgsino mass.  
Due to the $\left(1+\cos{\theta_{b\ell}}\right)$ factor, the matrix element peaks when the 
$b$-quark and the charged lepton are aligned. Later (Sec.~\ref{stRProDec.SEC}), we will point out that this $\cos\theta_{b\ell}$ behavior can be used to distinguish between $\stR\to b\ell\sNR$ and $t\to bW\to b\ell\nu$. It turns out that, for top quark decays, there is no peak in the event distribution when the $b$-quark and the charged lepton are aligned ({\it cf} Fig.~\ref{cthbE_1.FIG}).

The decay rate is  given by
\beq
\Gamma = 4 |Y_t|^2 |Y_N|^2 \frac{M_\stR^5}{M_{\tilde H}^4} \left[\frac{4\pi}{\left( 16 \pi^2 \right)^2} \hat{f}_{3PS} \right] \ , 
\label{stRGammaf.EQ}
\eeq
where $\hat{f}_{3PS}$ is a dimensionless 3-body phase space function. Note that $M_{\tilde H} > M_\stR$. 

Numerically, for $M_\sNR = 100~$GeV, %from Eq.~(\ref{stRGammaf.EQ}) 
%obtain the $\stR$ life-time to be 
\beq
c\tau \sim 10~{\rm mm}~\cdot \left(\frac{4\times 10^{-6}}{Y_N} \right)^2
\left(\frac{225~{\rm GeV}}{M_\stR} \right)^5 
\left( \frac{M_{\tilde H}}{250~{\rm GeV}} \right)^4 
\left( \frac{0.05}{\hat{f}_{3PS}} \right) \ . 
\label{stRdTvx.EQ}
\eeq
We remind the reader that $\hat{f}_{3PS}$ is a function of $M_\sNR, M_\stR$, and $M_{\tilde{H}}$.

\begin{figure}
\begin{center}
\scalebox{0.9}{\includegraphics{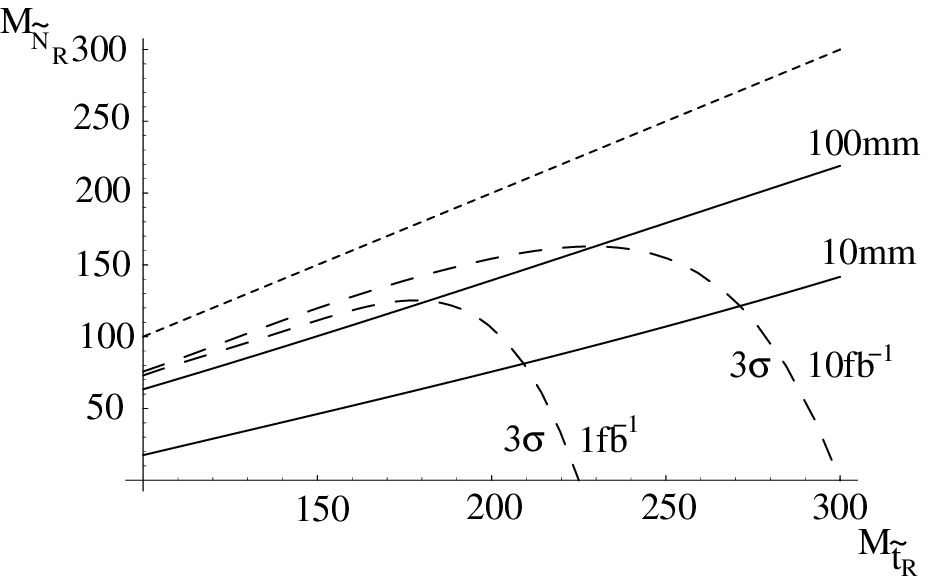}}\scalebox{0.9}{\includegraphics{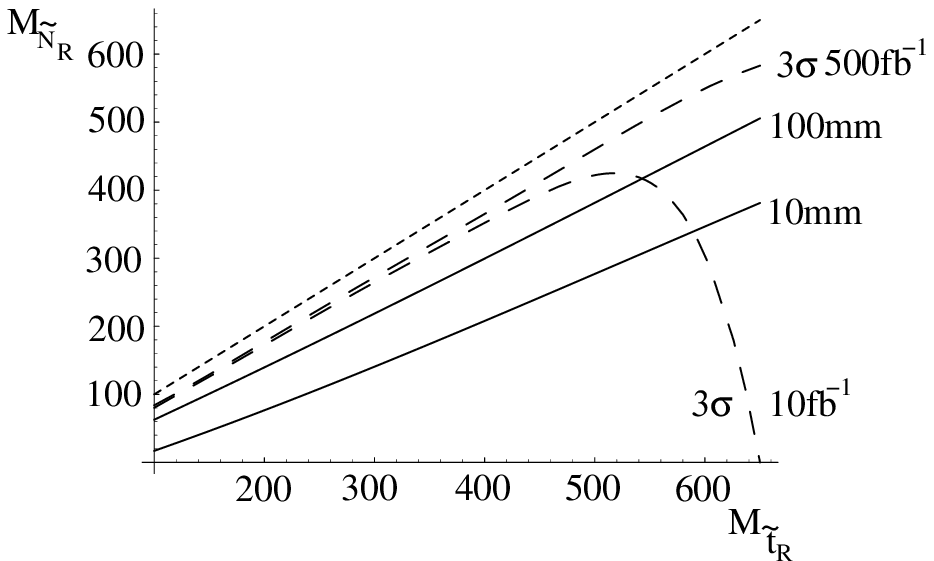}}
\caption{SOLID -- Contours of constant $c\tau$ for the right-handed stop, in the 
$M_\stR - M_\sNR$ plane. We assume $M_{\tilde{H}} = 1.1 M_\stR$. The $M_\stR = M_\sNR$ dotted line 
bounds the region of parameter space where $\sNR$ is the LSP. 
DASHED -- $3~\sigma$ sensitivity for stop pair production  at the Tevatron (left) and the 
LHC (right), for different integrated luminosities. Note that the displaced vertices are not 
taken into account when defining the sensitivity. See text for the details.  
\label{dvx_sig.FIG}}
\end{center}
\end{figure}

For $Y_N\sim 10^{-6}$, $c\tau$ can vary between a few millimeters to several meters, depending on the stop, sneutrino, and Higgsino masses. $c\tau$ values in the $M_\stR- M_\sNR$ plane are depicted in Fig.~\ref{dvx_sig.FIG}, assuming that $M_{\tilde{H}}=1.1M_\stR$. For larger values of the Higgsino mass, the constant $c\tau$ contours move toward the $M_\stR$-axis. Displaced vertices, {\it i.e.}, $c\tau\gtrsim 1$~mm are to be expected even for large scalar top masses, while for light enough stops and heavy enough LSPs and Higgsinos the stop may even be collider-stable. 
We elaborate on these possibilities in the next section. 

If lepton number is conserved in nature, $M_N=b_N=c_\ell=0$ (see App.~\ref{Theory.SEC}), neutrinos are Dirac fermions
and, in order to obtain the right order of magnitude for the active neutrino masses, $Y_N\sim 10^{-12}$ values are required 
({\it cf} Eq.~(\ref{mnuss.EQ})). In this case, $c\tau$ is measured in kilometers and, as far as collider phenomenology is concerned, the $\stR$ is absolutely stable. 

Heavy, strongly interacting, collider-stable SUSY particles are expected to  form 
`R-hadrons,' which behave like heavy nucleon-like objects. Similar experimental 
signatures have been studied elsewhere~\cite{Kraan:2004tz}.  
We do not pursue other experimental signatures of heavy, long-lived, hadronic states. 
We would, however, like to comment on the fact that very long-lived $\stR$ could form narrow 
``onium-like'' $\stR\stR^*$ bound states\footnote{The same would also be true of long-lived, 
gluino bound states~\cite{gluinonium}.}~\cite{super-onium}, whose decay 
may lead to a peak above the continuum background. 
An investigation of this phenomenon will be left for future work.

%%%%%%%%%%%%%%%%%%%%%%%%%%%%%%%%%%%%%%%%%%%%%%%%%%%%%%%%%%%%%%%%%%%%%%%%%%%%%%%%%%%%%%%%
\setcounter{footnote}{0}
\section{$\stR$ at the Tevatron and LHC}
\label{stRProDec.SEC}
We consider the production and decay of a $\stR$ NLSP at the Tevatron and the LHC followed by its 
subsequent decay into a purely right-handed sneutrino $\sNR$ LSP, {\it i.e.}, 
$\stR \rightarrow b \ell^+ \sNR$. Unless noted otherwise, we assume that the $\stR$ decays in this way 100\% of the time.
To illustrate the behavior of various observables, we will consider the case
$M_{\stR} = 225~$GeV, $M_\sNR = 100~$GeV, $M_{\sH} = 250~$GeV, $Y_N = 4\times 10^{-6}$.
In this case, $c\tau \approx 10~$mm  (see Eq.~(\ref{stRdTvx.EQ})).

\begin{figure}
\begin{center}
\scalebox{0.5}{\includegraphics{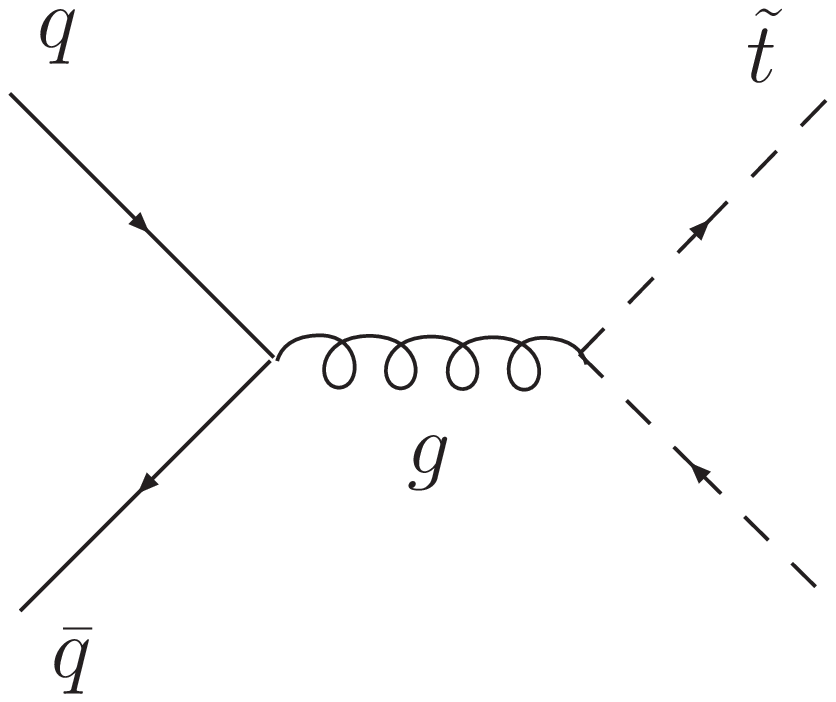}}\hspace*{0.25cm}
\scalebox{0.5}{\includegraphics{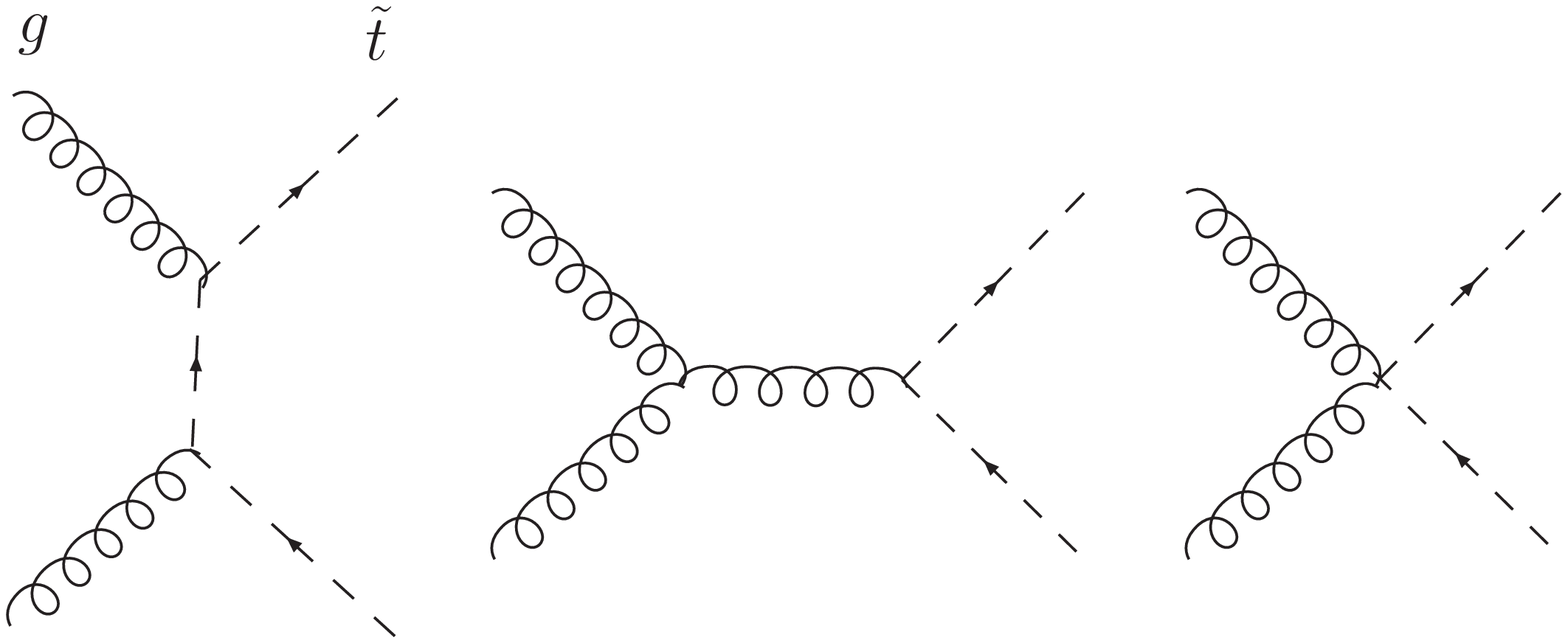}}
\caption{
Dominant parton level $\tilde t$ pair-production processes at a hadron collider. 
The diagrams with gluons in the initial state dominate at the LHC.
\label{pp2stop.FIG}}
\end{center}
\end{figure}

Fig.~\ref{pp2stop.FIG} depicts the dominant production mechanism of a $\stR$ pair\footnote{
Depending on the details of the SUSY spectrum, there may be other $\stR$ production channels, including gluino pair production, followed by $\tilde{g}\to \bar{t}\stR$ (if kinematically accessible). Here we concentrate on ``direct'' QCD production.}, at leading order.
The dominant contribution to  $\stR$ decay is depicted in Fig.~\ref{sTdecay.FIG}. The
$\sNR$ escapes the detector as missing energy so that, experimentally, one should observe
 $p p (\bar{p}) \rightarrow \stR \stR^*
\rightarrow b \ell^+ \bar{b} \ell^- + \mEt$. The dominant physics background
for this process is expected to be top quark pair production: $p p (\bar{p}) \rightarrow t
\bar t \rightarrow b W^+ \bar{b} W^- \rightarrow b \ell^+ \bar{b}
\ell^- + \mEt$, where the missing energy is due to
neutrinos in the final state.
We use the Monte Carlo program Pythia (version 6.327) in order to analyze,
at the parton level, the signal and background at the Tevatron and the LHC.

The observables available in order to discriminate the signal from
background are the transverse components of ${\bf k}_b$, ${\bf k_{\ell^+}}$,
${\bf k}_{\bar b}$ and ${\bf k_{\ell^-}}$. We impose the ``level~1'' cuts
shown in Table~\ref{level1.TAB}, necessitated by the detector geometric acceptance and
thresholds. For the isolation cut, we make the standard definition 
$R_{b\ell}^2 \equiv (\phi_b - \phi_\ell)^2 + (\eta_b - \eta_\ell)^2$.
These values are typical of other Tevatron analyses, and are meant to 
approximate the capabilities of the LHC detectors.
\begin{table}
\caption{``Level~1'' cuts imposed in our analysis. We define 
$R_{b\ell}^2 \equiv (\phi_b - \phi_\ell)^2 + (\eta_b - \eta_\ell)^2$.
\label{level1.TAB}}
\begin{tabular}{|c||c|c|}
\hline 
1.&
Rapidity cuts&
$|\eta_{\ell}|<2.5$, $|\eta_{b}|<2.5$\tabularnewline
\hline 
2.&
$p_{T}$ cuts&
${p_{T}}_{\ell}>20~$GeV, ${p_{T}}_{b}>10~$GeV\tabularnewline
\hline 
3.&
Isolation cut&
$R_{b\ell}>0.4$\tabularnewline
\hline
\end{tabular}
\end{table}

\begin{figure}
\begin{center}
\scalebox{0.45}{\includegraphics[angle=270]{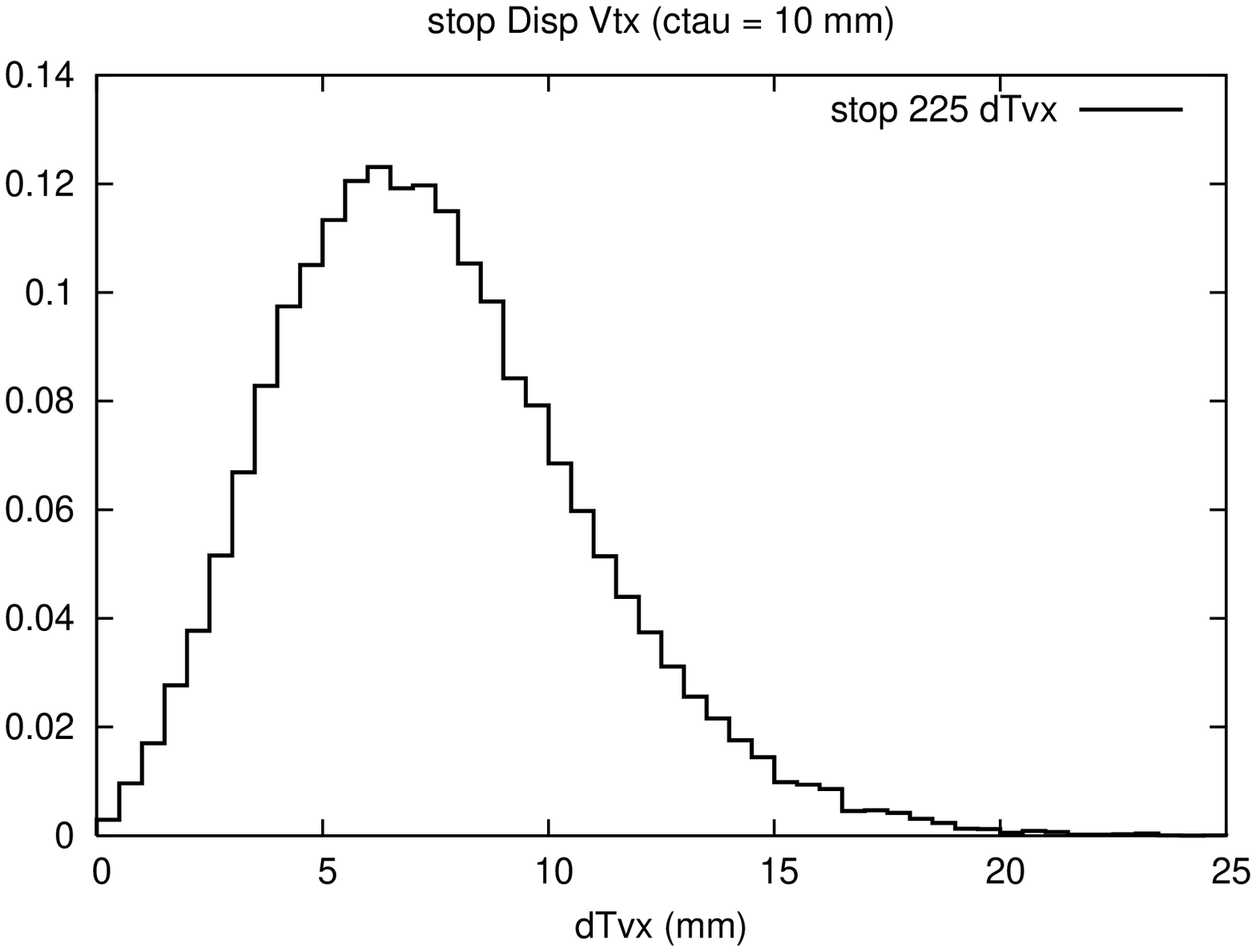}}\scalebox{0.45}{\includegraphics[angle=270]{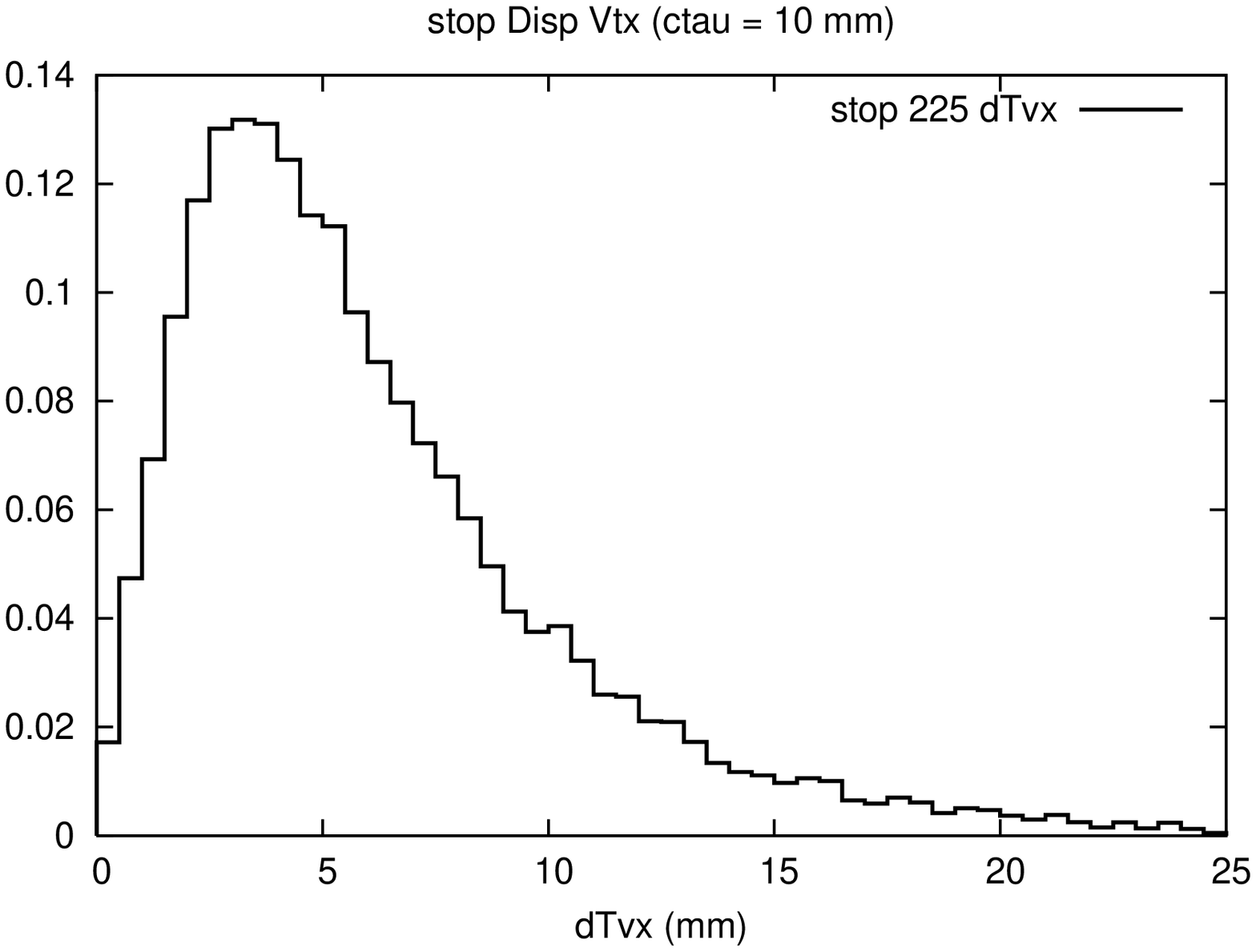}}
\caption{
The distributions of the transverse displacement of the stop (in mm), at the Tevatron (left) and the 
LHC (right).  
\label{dTvx_1.FIG}}
\end{center}
\end{figure}
Fig.~\ref{dTvx_1.FIG} depicts the resulting
transverse displacement (in the $x-y$ plane), after including the boost
of the stop at the Tevatron (left) and the LHC (right).
Even though
the stop is more boosted at the LHC (compared to the Tevatron) due to
the higher center-of-mass energy, the stop pair is more forward peaked
at the LHC, causing their transverse displacement to be smaller at the
LHC (compared to the Tevatron). For other $M_\sNR$ and $M_\stR$ values, the
 transverse displacement scales with the scalar top's $c\tau$. These have been 
 discussed in the previous section, and are depicted in Fig.~\ref{dvx_sig.FIG}.

If a stop is long-lived enough, it will behave like a stable or quasi-stable hadronic object (as commented on earlier).  
If it decays before exiting the tracking subsystem, a displaced
vertex may be reconstructed through the stop decay products'
3-momenta meeting away from the primary interaction point.  
In the example considered, a displaced vertex of $10~$mm can be easily
discerned at the Tevatron and the LHC.
On each side, the $b$-quark itself leads to an additional
displaced vertex, and its 3-momentum vector can be reconstructed from
its decay products. In combination with the 3-momentum of the
lepton, the stop displaced vertex can be determined.  In order to
reveal the displaced vertex, one must require either the $b$-quark or the charged lepton
3-momentum vector to miss the primary vertex.  Since a pair of stops
is produced, we would expect to discern two displaced vertices in the
event (not counting the displaced vertices due to the b-quarks).  If a
$b$-quark ($\ell^+$)  cannot be distinguished from a $\bar b$-antiquark ($\ell^-$), the
associated combinatoric problem of assigning the decay products to the
parent stop will have to be considered. Such an event with two
displaced vertices, from each of which originates a high $p_T$ $\ell$ and 
$b$-quark, is quite uncommon in SUSY models\footnote{Other SUSY models
that lead to a displaced vertex are the NMSSM with a singlino 
LSP~\cite{Hesselbach:2000qw},
the case of bilinear $R$-parity breaking where neutrino masses are generated
via $R$-parity breaking effects~\cite{deCampos:2005ri}, and
also those in \cite{DispVtxRefs}.} and might prove
to be one of the main distinguishing characteristics of such a
scenario. A cut on the displaced vertex provides for a very
effective way to separate stop events from the top background. If one can efficiently 
explore such cuts, we anticipate that NLSP scalar top searches may turn out to be 
physics-background free.

Another characteristic feature is the non-universal rates for decays into 
$e$, $\mu$ and $\tau$ leptons. This is to be expected given that the stop decays are
proportional to the $Y_N$'s which are in general different for the three leptons. Here, we do not 
take advantage of this feature. 

If the stop displaced vertex cannot  be efficiently resolved,
one will have to resort to  more conventional analysis methods. In the remainder of this section,
we explore various kinematical distributions for both the signal (right-handed scalar top pair production) and the physics background (top pair production), obtained after imposing the level~1 cuts
listed in Table~\ref{level1.TAB}. 
All the distributions are in the lab frame, normalized to unit area. Note that the analysis performed here
also applies to other SUSY scenarios in which the scalar top decays predominantly to bottom plus charged lepton plus missing energy, regardless of whether the stop decays promptly or leaves behind a displaced vertex.

\begin{figure}
\begin{center}
\scalebox{0.45}{\includegraphics[angle=270]{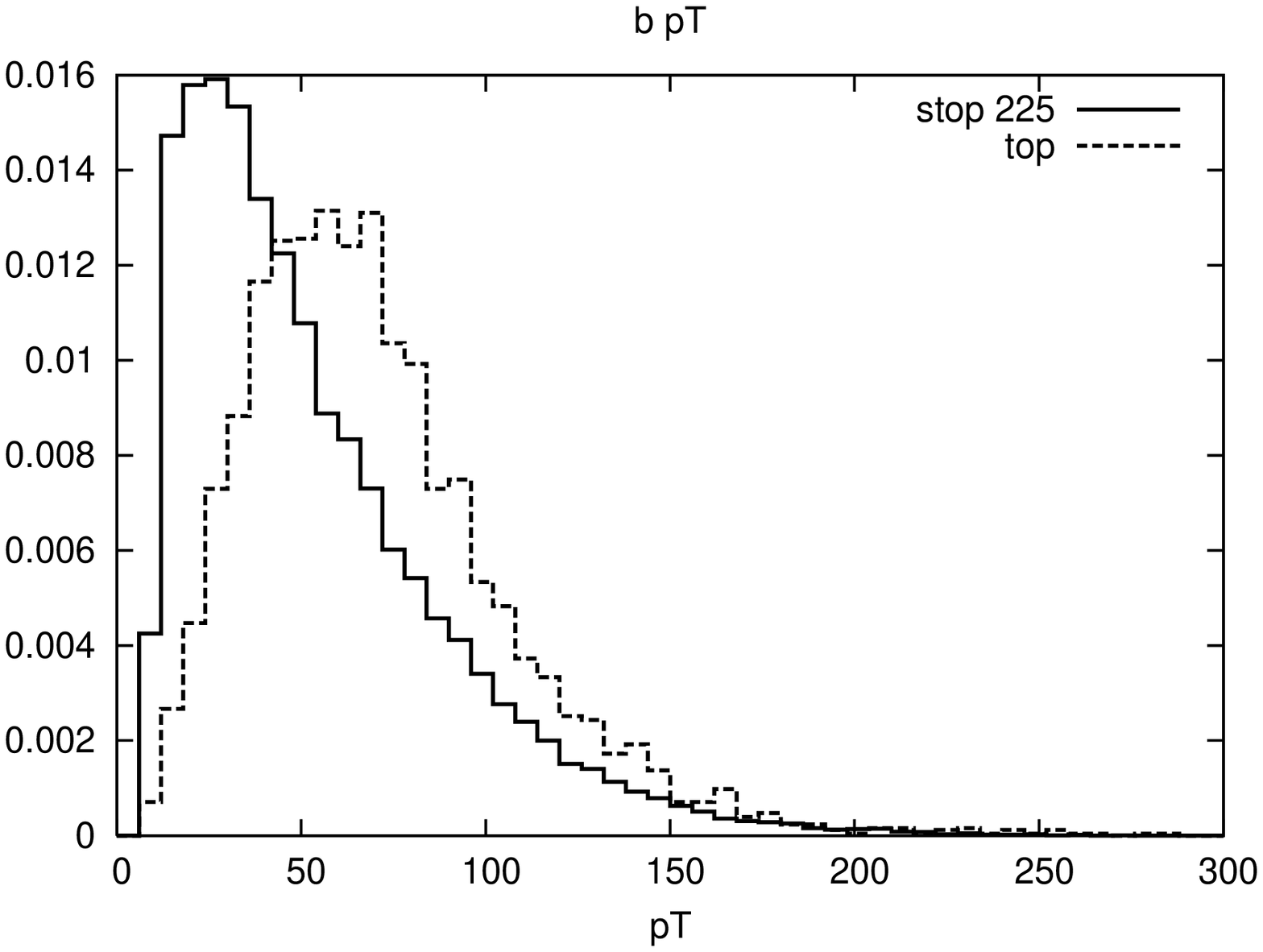}}\scalebox{0.45}{\includegraphics[angle=270]{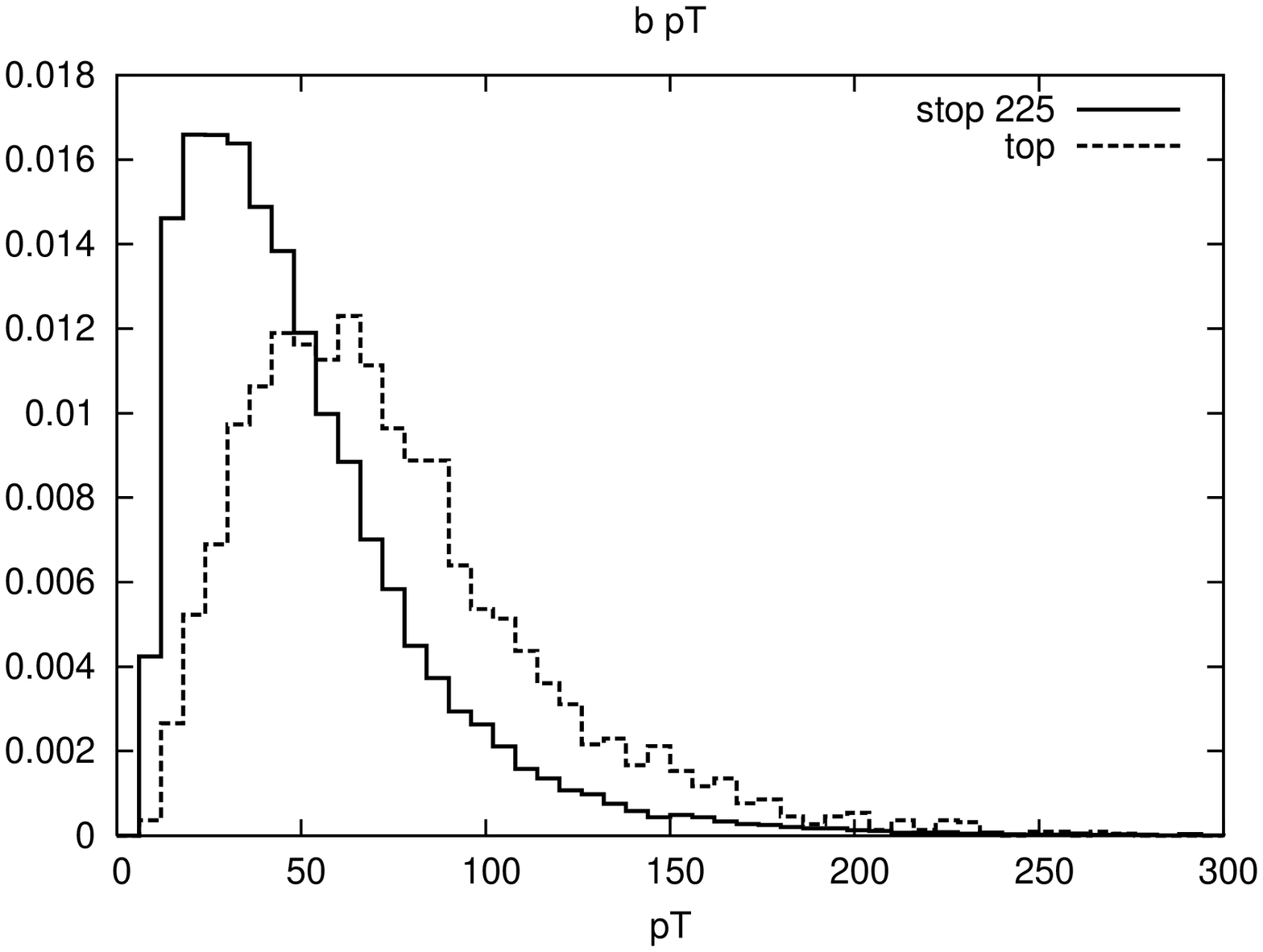}}
\caption{
The distribution of $p_T$ of the $b$-quark at the Tevatron (left) and LHC (right) resulting from the 
decay of a 225~GeV stop and a top quark.
\label{pTb_1.FIG}}
\end{center}
\end{figure}
Fig.~\ref{pTb_1.FIG} depicts the parton level distributions of the $p_T$ of the $b$-quark  
at the Tevatron (left) and the LHC (right) resulting from a 225~GeV stop and from a top, 
obtained using Pythia.
\begin{figure}
\begin{center}
\scalebox{0.45}{\includegraphics[angle=270]{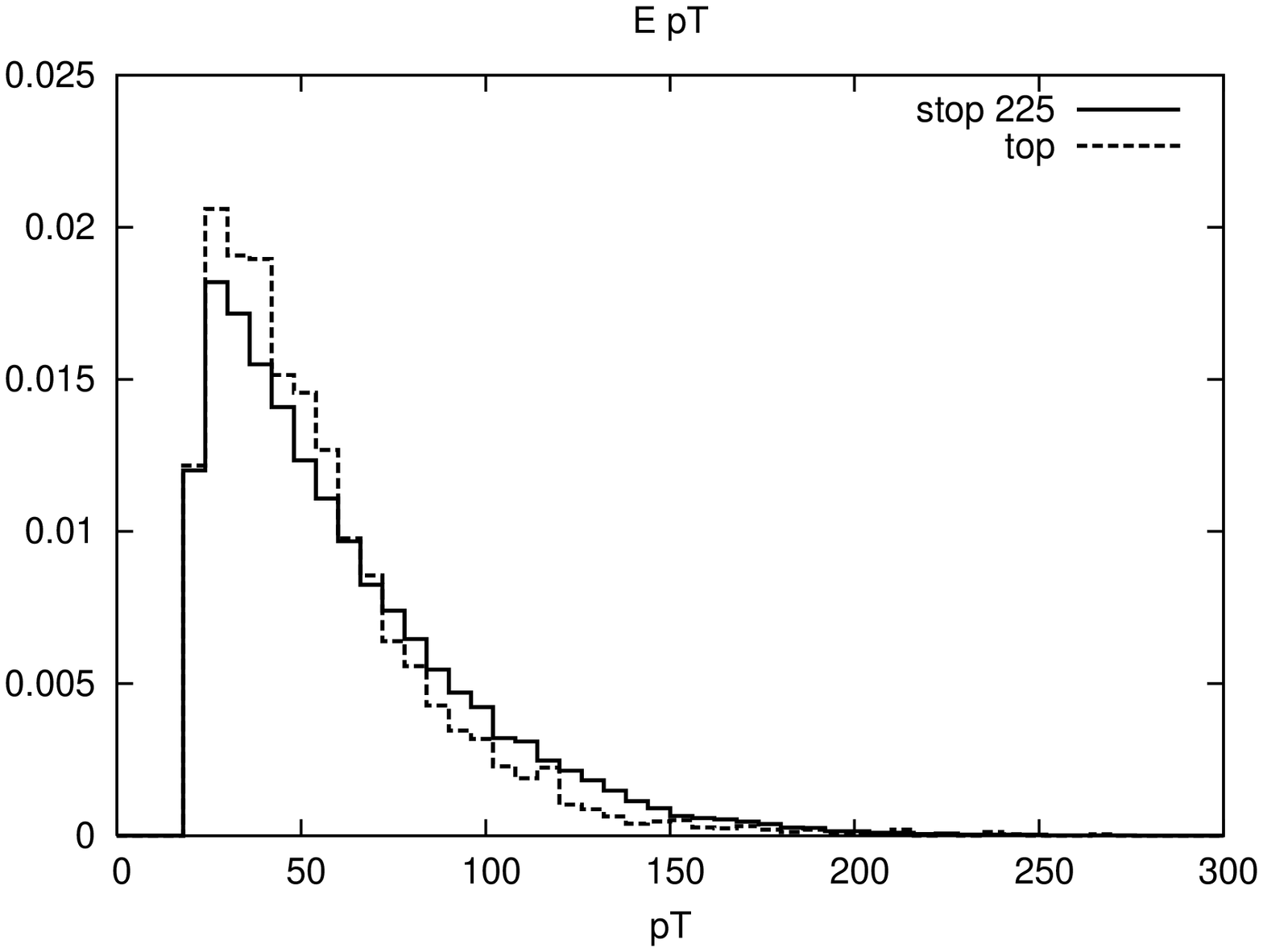}}\scalebox{0.45}{\includegraphics[angle=270]{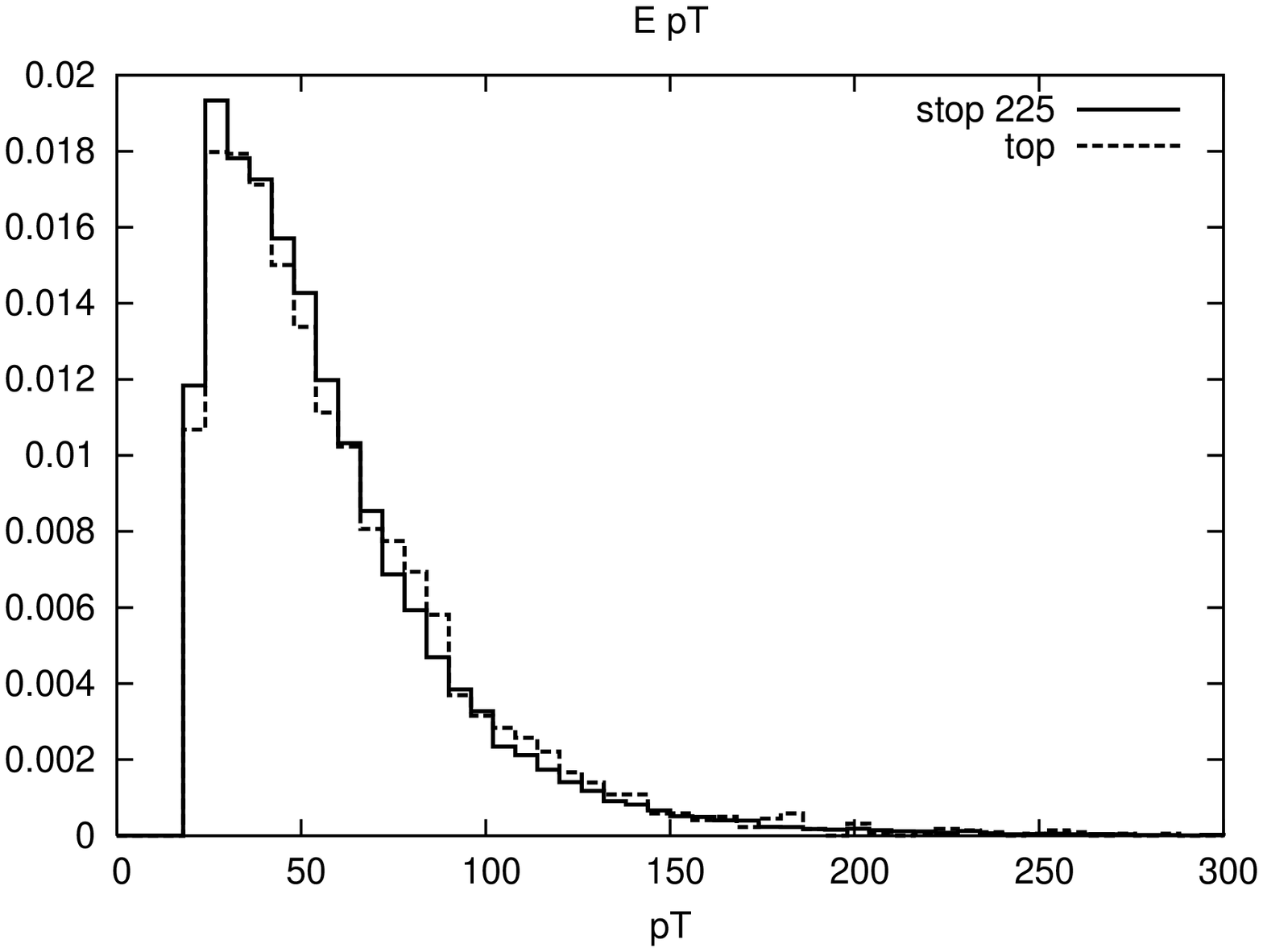}}
\caption{
The distribution of $p_T$ of the charged lepton resulting from the decay of a 
225~GeV stop, and a top, at the Tevatron (left) and the LHC (right).
\label{pTE_1.FIG}}
\end{center}
\end{figure}
Fig.~\ref{pTE_1.FIG} depicts the distributions of the $p_T$ of the charged lepton at the Tevatron (left) 
and the LHC (right), resulting from a 225~GeV stop, and from a top.
The $p_T$ of the $b$-quark from the $225~$GeV stop peaks at a lower value compared 
to the top quark background, and therefore accepting them at high efficiency for 
$p_T \lesssim 40~$GeV will be very helpful in maximizing the signal acceptance.
The signal and background shapes are quite similar and no simple set of $p_T$ cuts can be 
made in order to significantly separate signal from background. 
As the mass difference between the $\stR$ and the $\sNR$ decreases, the $b$-quark and the charged lepton $p_T$ distributions peak at lower values due to less available phase-space, making the measurements more 
challenging.

\begin{figure}
\begin{center}
\scalebox{0.45}{\includegraphics[angle=270]{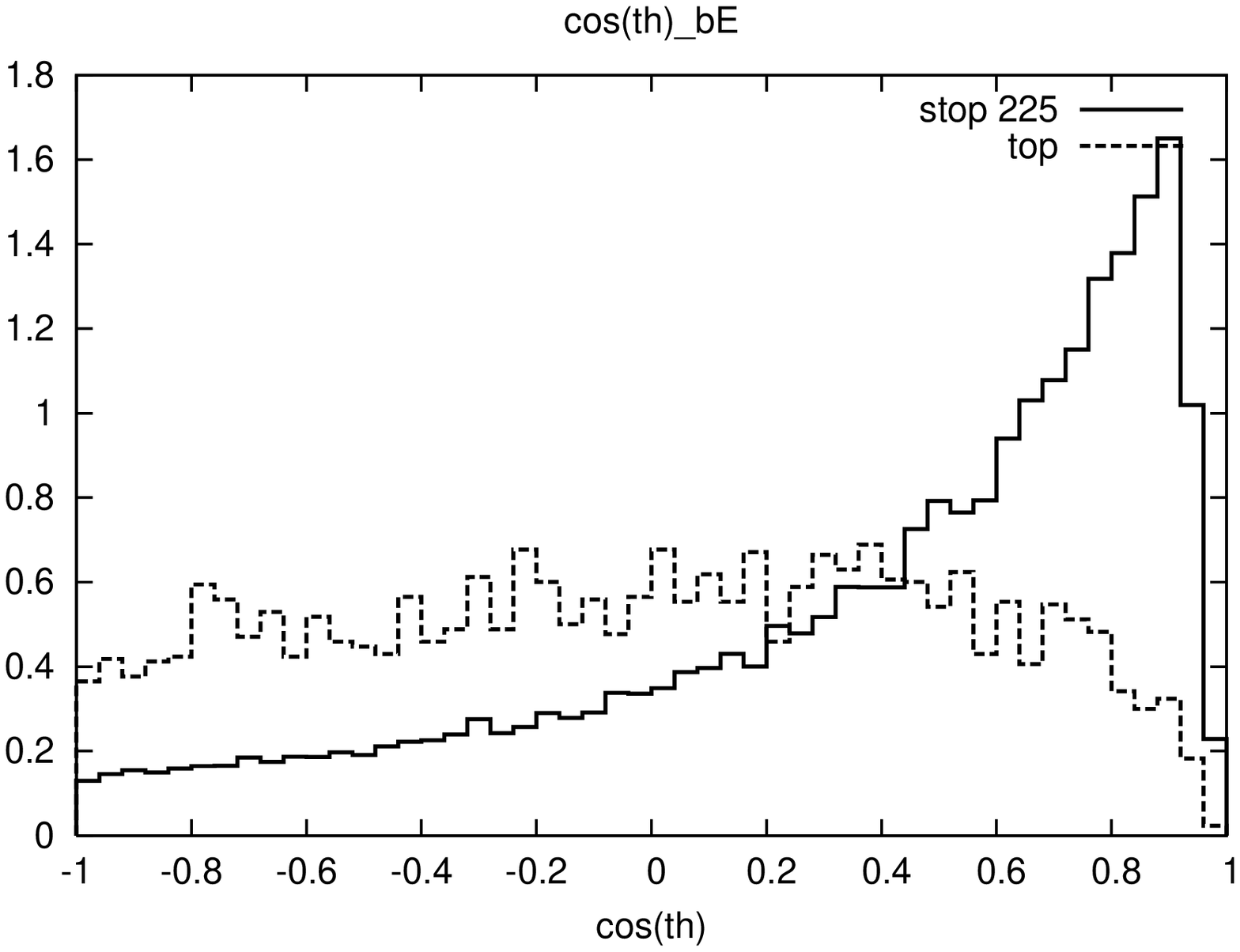}}\scalebox{0.45}{\includegraphics[angle=270]{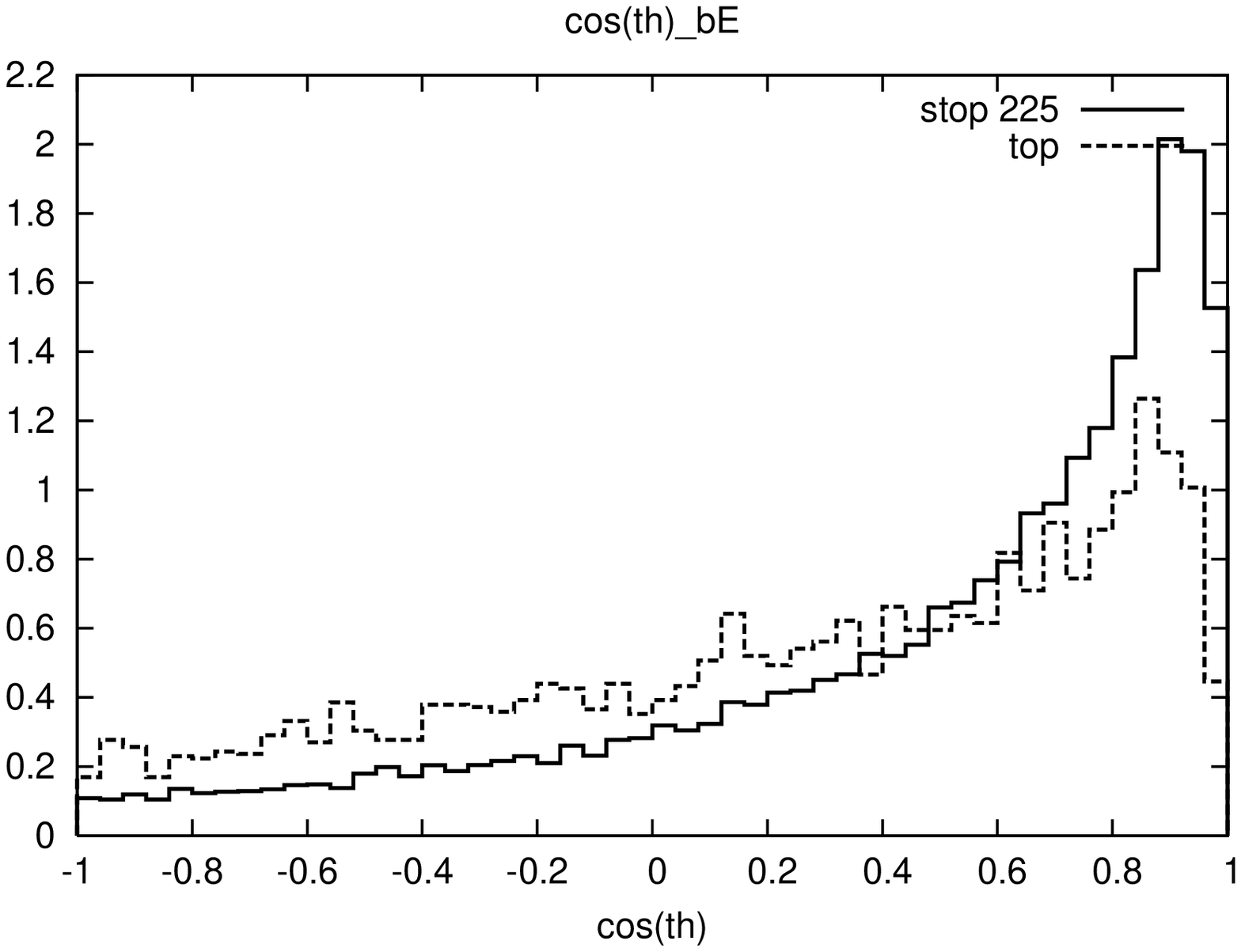}}
\caption{
The distribution of $\cos\theta_{b\ell}$ resulting from the decay of a 
225~GeV stop, and a top, at the Tevatron (left) and LHC (right).
\label{cthbE_1.FIG}}
\end{center}
\end{figure}
Fig.~\ref{cthbE_1.FIG} depicts the distribution of $\cos\theta_{b\ell}$, the angle between
the 3-momenta ${\bf k}_b$ and $\bf k_\ell$, for both the signal and background.
It is important to appreciate that, by default, Pythia generates stop decays into the 
3-body final state according only to phase-space,
ignoring the angular dependence of the decay matrix element. We have reweighted
Pythia events to include the correct angular dependence in the 
decay matrix element. 
Consistent with the expectation from Eq.~(\ref{TfistR.EQ}), we see for the signal that the 
distribution peaks for the $b$-quark and charged lepton 3-momenta aligned, unlike the 
background.\footnote{For another discussion on how to extract stop NLSP's (in a different context) 
see \cite{Chou:1999zb}.}  
It is unfortunate that  the isolation level~1 cut (see Table~\ref{level1.TAB}) on the 
leptons removes  more 
signal events than background events. Relaxing this constraint  as much as practical 
would help in this regard.

\begin{figure}
\begin{center}
\scalebox{0.45}{\includegraphics[angle=270]{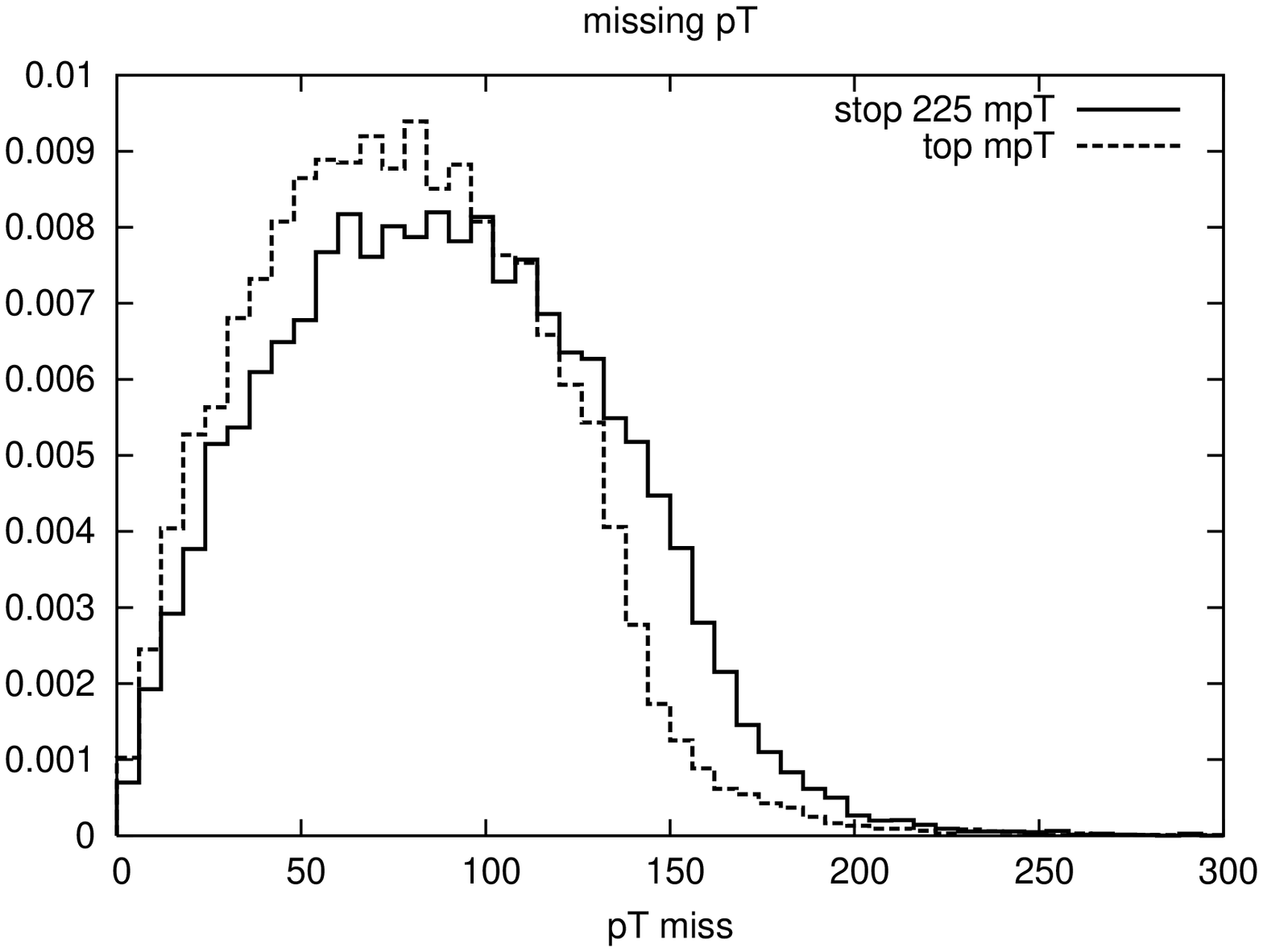}}\scalebox{0.45}{\includegraphics[angle=270]{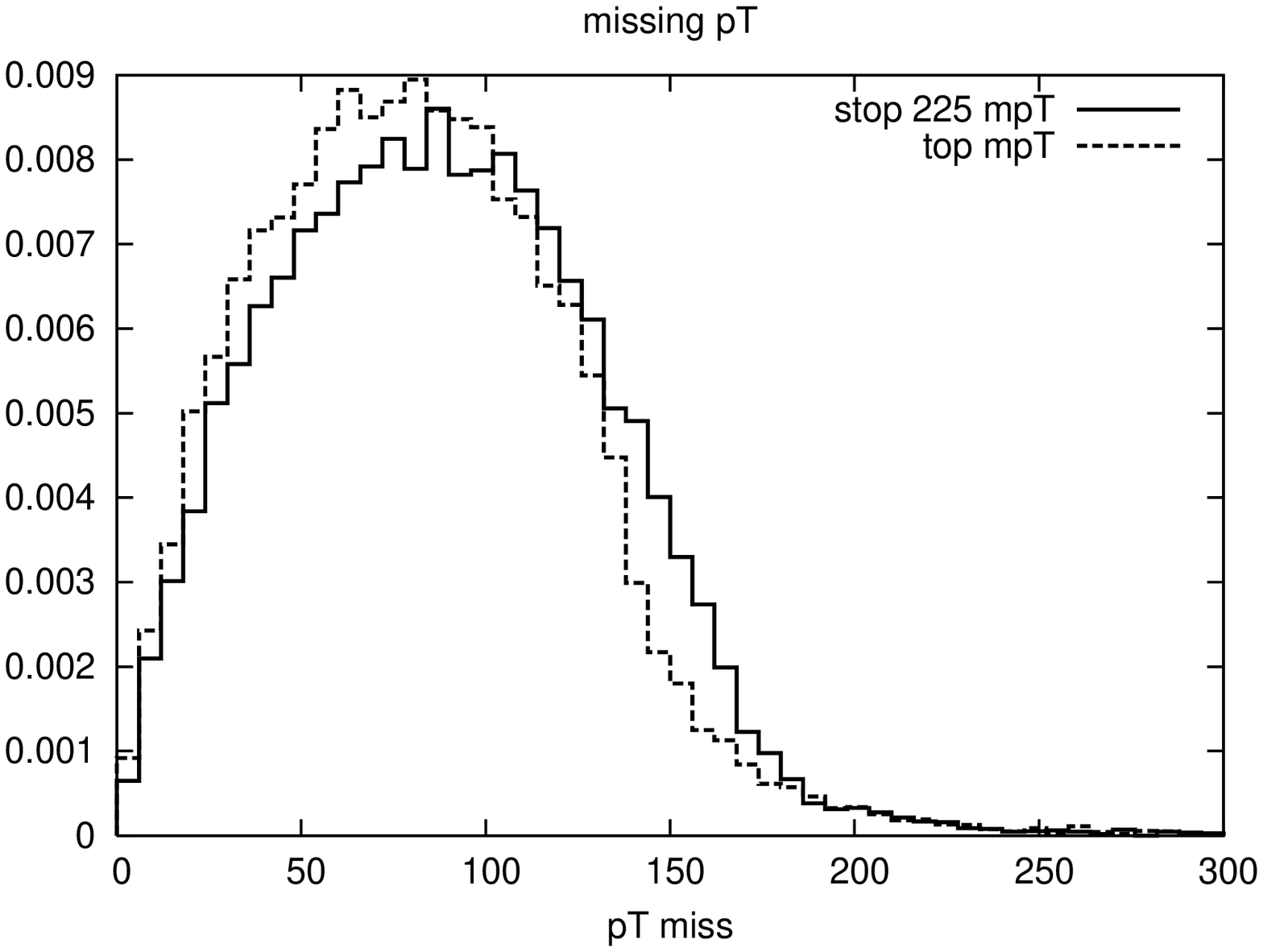}}
\caption{
The total $\mpt$ distribution resulting from a 225~GeV stop, and a top, 
at the Tevatron (left) and the LHC (right).
\label{misspT_1.FIG}}
\end{center}
\end{figure}
Fig.~\ref{misspT_1.FIG} shows the total $\mpt$ distribution resulting
from the production and decay of a 225~GeV stop and a top quark
for the Tevatron (left) and the LHC (right). Here $\mpt$ is defined as
$\sqrt{{\mpt}_x^2 + {\mpt}_y^2}$, with ${\mpt}_x$ and ${\mpt}_y$ the
$x$ and $y$ components of the total missing momentum vector.  Pythia
ignores the spin correlation between the two opposite side particles and
therefore the $\mpt$ distributions shown are not entirely accurate in
case of the $t$-quarks. 
However, spin correlation modifies the $\mpt$ distribution to only
a small degree~\cite{Barger:1988jj}, although it can lead to significant 
effects in suitably chosen observables~\cite{SpinCorrRefs}.

The angular correlation of the stop pair is different from that of the top quark pair,
since the former is a scalar and the latter a fermion. 
We expect the quantities ${\bf k}_b\cdot{\bf k}_{\bar b}$ and 
$\bf k_{\ell^+}\cdot\bf k_{\ell^-}$ to inherit some of this 
difference, making these potentially good discriminants for signal and background. 
As just mentioned, Pythia is not suitable to investigate this aspect since it does not retain 
spin correlations, and we postpone this investigation to future work. 

\begin{table}
\caption{Signal (stop) and background (top) pair production cross-section
at the Tevatron, with $\epsilon_b =0.5$, and $\epsilon_\ell = 0.9$. 
$\sigma_S$ and $\sigma_B$ denote the signal and background cross-sections, 
$\alpha$ the fraction that passes  level~1 cuts (see Table~\ref{level1.TAB}), $S$ and $B$ the 
number of signal and background events for $1~{\rm fb}^{-1}$.
\label{TevReach.TAB}}
\begin{tabular}{|c|c|c|c|c|c|c|c|c|c|}
\hline 
$M_\stR$&
$M_\sNR$&
$\sigma_{S}$(pb)&
$\sigma_{B}$(pb)&
$\alpha$&
$S$&
$B$&
$S/B$&
$S/\sqrt{B}$&
$S/\sqrt{S+B}$\tabularnewline
\hline
\hline 
\multirow{2}{0.75cm}{100}&
50&
11.83&
6.77&
0.26&
162&
9&
18.93&
55.36&
12.4\tabularnewline
\cline{2-2} \cline{3-3} \cline{4-4} \cline{5-5} \cline{6-6} \cline{7-7} \cline{8-8} \cline{9-9} \cline{10-10} 
\multicolumn{1}{|c|}{}&
75&
11.83&
6.77&
0.04&
4&
9&
0.45&
1.31&
1.09\tabularnewline
\cline{1-1} 
\hline 
\multicolumn{1}{|c|}{\multirow{2}{0.75cm}{150}}&
100&
1.24&
6.77&
0.29&
21&
9&
2.46&
7.21&
3.87\tabularnewline
\cline{2-2} \cline{3-3} \cline{4-4} \cline{5-5} \cline{6-6} \cline{7-7} \cline{8-8} \cline{9-9} \cline{10-10} 
&
125&
1.24&
6.77&
0.05&
1&
9&
0.07&
0.21&
0.21\tabularnewline
\hline 
\multicolumn{1}{|c|}{\multirow{2}{0.75cm}{175}}&
100&
0.48&
6.77&
0.47&
22&
9&
2.53&
7.39&
3.93\tabularnewline
\cline{2-2} \cline{3-3} \cline{4-4} \cline{5-5} \cline{6-6} \cline{7-7} \cline{8-8} \cline{9-9} \cline{10-10} 
&
150&
0.48&
6.77&
0.05&
0.2&
9&
0.03&
0.08&
0.08\tabularnewline
\hline 
\multicolumn{1}{|c|}{\multirow{2}{0.75cm}{250}}&
100&
0.04&
6.77&
0.71&
4&
9&
0.48&
1.4&
1.15\tabularnewline
\cline{2-2} \cline{3-3} \cline{4-4} \cline{5-5} \cline{6-6} \cline{7-7} \cline{8-8} \cline{9-9} \cline{10-10} 
&
200&
0.04&
6.77&
0.31&
1&
9&
0.09&
0.27&
0.26\tabularnewline
\hline
\end{tabular}
\end{table}

\begin{table}
\caption{Signal (stop) and background (top) pair production cross-section
at the LHC, with $\epsilon_b =0.5$, and $\epsilon_\ell = 0.9$. 
$\sigma_S$ and $\sigma_B$ denote the signal and background cross-sections, 
$\alpha$ the fraction that passes level~1 cuts (see Table~\ref{level1.TAB}), $S$ and $B$ the 
number of signal and background events for $10~{\rm fb}^{-1}$.
\label{LHCReach.TAB}}

\begin{tabular}{|c|c|c|c|c|c|c|c|c|c|}
\hline 
$M_\stR$&
$M_\sNR$&
$\sigma_{S}$(pb)&
$\sigma_{B}$(pb)&
$\alpha$&
$S$&
$B$&
$S/B$&
$S/\sqrt{B}$&
$S/\sqrt{S+B}$\tabularnewline
\hline
\hline 
\multirow{2}{0.75cm}{100}&
75&
1332.5&
873&
0.03&
1938&
8662&
0.22&
20.82&
18.82\tabularnewline
\cline{2-2} \cline{3-3} \cline{4-4} \cline{5-5} \cline{6-6} \cline{7-7} \cline{8-8} \cline{9-9} \cline{10-10} 
\multicolumn{1}{|c|}{}&
83&
1332.5&
873&
0.01&
73&
8662&
0.01&
0.78&
0.78\tabularnewline
\cline{1-1} 
\hline 
\multicolumn{1}{|c|}{\multirow{2}{0.75cm}{150}}&
100&
228.79&
873&
0.23&
25325&
8662&
2.92&
272.1&
137.37\tabularnewline
\cline{2-2} \cline{3-3} \cline{4-4} \cline{5-5} \cline{6-6} \cline{7-7} \cline{8-8} \cline{9-9} \cline{10-10} 
&
128&
228.79&
873&
0.02&
144&
8662&
0.02&
1.54&
1.53\tabularnewline
\hline 
\multicolumn{1}{|c|}{\multirow{2}{0.75cm}{250}}&
200&
21.32&
873&
0.26&
2886&
8662&
0.33&
31.01&
26.86\tabularnewline
\cline{2-2} \cline{3-3} \cline{4-4} \cline{5-5} \cline{6-6} \cline{7-7} \cline{8-8} \cline{9-9} \cline{10-10} 
&
225&
21.32&
873&
0.03&
40&
8662&
0.01&
0.43&
0.43\tabularnewline
\hline 
\multicolumn{1}{|c|}{\multirow{2}{0.75cm}{500}}&
400&
0.56&
873&
0.59&
398&
8662&
0.05&
4.28&
4.19\tabularnewline
\cline{2-2} \cline{3-3} \cline{4-4} \cline{5-5} \cline{6-6} \cline{7-7} \cline{8-8} \cline{9-9} \cline{10-10} 
&
425&
0.56&
873&
0.48&
263&
8662&
0.03&
2.83&
2.78\tabularnewline
\hline
\multirow{2}{0.75cm}{650}&
250&
0.14&
873&
0.83&
195&
8662&
0.02&
2.10&
2.08\tabularnewline
%\hline
\cline{2-2} \cline{3-3} \cline{4-4} \cline{5-5} \cline{6-6} \cline{7-7} \cline{8-8} \cline{9-9} \cline{10-10}
&
500&
0.14&
873&
0.71&
145&
8662&
0.02&
1.56&
1.54\tabularnewline
\hline
\end{tabular}
\end{table}

The sensitivity of the Tevatron and the LHC to the stop NLSP is depicted in 
Table~\ref{TevReach.TAB} and Table~\ref{LHCReach.TAB}, respectively,  for various 
$\stR$ and $\sNR$ masses.
The stop pair-production cross-section from Pythia multiplied by the appropriate 
K-factor~\cite{Beenakker:1997ut} is displayed in the table for the Tevatron and the LHC, 
along with the top pair-production cross-section~\cite{Kidonakis:2004hr}. 
The fraction of signal events that passes level~1 cuts (specified in Table~\ref{level1.TAB}), 
denoted $\alpha$, obtained using Pythia
is shown, and the fraction of background events (not shown in the table) 
is 0.79 (Tevatron), and 0.7 (LHC).  
We compute the number of signal events ($S$), background events ($B$), for 
$1~{\rm fb}^{-1}$ (Tevatron) and $10~{\rm fb}^{-1}$ (LHC), and compute the figures-of-merit
$S/B$, $S/\sqrt{B}$ and $S/\sqrt{S+B}$. 
We have taken the b-tagging efficiency $\epsilon_b = 0.5$ and the lepton 
identification efficiency $\epsilon_\ell = 0.9$. 
The number of events is given by $\sigma\, {\cal L}\, \alpha^2\, \epsilon_b^2\, \epsilon_\ell^2$,
where ${\cal L}$ is the integrated luminosity. 
It is important to emphasize that we do not
consider the possibility that one can efficiently identify that the scalar top decays far from the production point. If this is the case, we anticipate that physics backgrounds will be significantly reduced. At the same time, we remind the reader that we have fixed $M_{\tilde{H}}/M_\stR=1.1$. For larger values of this ratio, scalar tops are expected to be (much) longer-lived, in which case other approaches to data analysis are required. 

Fig.~\ref{dvx_sig.FIG} shows the projected $3\, \sigma$ contours in the 
$M_\stR-M_\sNR$ plane, with $1~{\rm and}~10~{\rm fb}^{-1}$ at the 
Tevatron\footnote{The $\tilde t \rightarrow b \ell \snuL$ mode at the Tevatron
has been studied in the $j\ell^+\ell^- \mEt$ channel in \cite{Abel:2000vs}
with qualitatively similar results to ours.}, and
$10~{\rm and}~500~{\rm fb}^{-1}$ at the LHC. 
We find from our analysis with just level~1 cuts (see Table~\ref{level1.TAB}) that the Tevatron can probe stop
masses up to about 300~GeV (with 10~${\rm fb}^{-1}$), while the LHC
reach extends slightly above 650~GeV (with 500~${\rm fb}^{-1}$), with the
reach depending on the $\sNR$ mass.
We expect that the sensitivity can be improved  with more sophisticated cuts 
and a multivariate analysis.
A smaller $M_\stR-M_\sNR$ mass difference leads to softer $b$-quarks and charged leptons, 
resulting in a smaller $\alpha$ and leading to a lower statistical significance.
Here we assume that the stop decays to the leptonic channel considered with 
100~\% branching ratio. In a specific model the actual significance can be obtained by
including the branching ratio. 
Our estimates for the top and 
stop production rates are in good qualitative agreement with experimental 
measurements of top production~\cite{top_exptRef} 
and bounds imposed by stop searches at the Tevatron~\cite{stop_exptRef}.

%%%%%%%%%%%%%%%%%%%%%%%%%%%%%%%%%%%%%%%%%%%%%%%%%%%
\section{Production and decay of other SUSY particles}
\label{OthSusy.SEC}
In this section we briefly discuss some aspects
of unique signatures associated with the production and decay of other SUSY particles.
The actual signatures depend on the detailed SUSY spectrum. In particular, we digress
on features associated with other NLSP candidates, including sbottoms, gluinos, gauginos 
and sleptons. We also comment on the fate of the ``other'' mostly right-handed sneutrinos, which
are expected to be, as far as collider experiments are concerned, also stable.

%%%%%%%%%%%%%%%%%%%%%%%%%%%%%%%
\subsection{sbottom}
If the $\sbR$ is the NLSP, its pair-production cross-section is naively similar to the stop NLSP 
case discussed
in Sec.~\ref{stRProDec.SEC}. The decay channel analogous to the one discussed for the stop NLSP is $\tilde{b}\to t\ell\sNR$. This decay channel, however, may not be kinematically accessible.
Regardless, the dominant channel for a large chunk of the parameter space is expected to be
be $\sbR \rightarrow b \nu \sNR$, resulting in the signature 
$p p(\bar{p}) \rightarrow b \bar{b} + \mEt$. 
Due to the additional suppression of the decay rate by $Y_b^2$, we expect NLSP sbottoms
to be more long-lived than the stop NLSP 
by a factor of $Y_t^2/Y_b^2$.
Here, we expect that it will be harder (compared to the $\stR$ NLSP case) to identify efficiently
a potentially displaced vertex, due to the absence of charged leptons in the final state.
One may be able to achieve this by asking whether the 
 reconstructed 3-momenta of the opposite side $b$-quarks point back to the primary vertex.
One source of background is QCD $b \bar{b}$ production -- huge -- 
but demanding substantial $\mEt$ should help extract the signal.
We expect that substantial $\mEt$ will also be crucial for triggering on the event. 

The decay mode $\sbR \rightarrow c \ell \sNR^*$ is suppressed by $|V_{cb}|^2$ compared to
the mode discussed above, and should have a branching ratio around $10^{-3}$. This mode 
leads to the signature $p p(\bar{p}) \rightarrow c \bar{c} \ell^+ \ell^- + \mEt$. The additional
pair of leptons will help in discriminating these events from background but, 
given the small branching ratio,
we expect this decay mode to be out of the reach of the Tevatron (but not the LHC).  
As with stop NLSPs, given that the decay is proportional to the $Y_N$'s, we generically expect different rates for different charged lepton final states  as a characteristic of this scenario.

%%%%%%%%%%%%%%%%%%%%%%%%%%%%%%%
\subsection{Gluino}
If the gluino is the NLSP, it decays primarily to a four-body final state
$\tilde g \rightarrow q \bar{q}^\prime \ell^+ \sNR$, 
via off-shell $\tilde q^\prime$ and $\schi^+$. 
This leads to the signature $p p (\bar{p}) \rightarrow 4j + 2 \ell^\pm + \mEt$.
A related mode is to a neutrino in the final state, i.e., 
$\tilde g \rightarrow q \bar{q} \bar\nu \sNR$ via off-shell $\tilde q$ and $\schi^0$,
leading to the signature $p p (\bar{p}) \rightarrow 4j + \mEt$. Unlike the previous mode, 
in this mode the leptons are unobservable.

Owing to the four-body final state and the  $Y_N$ suppression, the gluino is 
expected to be quite long lived. The decay rate is 
\bea
\Gamma_{\tilde g} \sim g_s^2 g^2 Y_N^2 \frac{M_{\tilde g}^7}{M_\chi^2 M_{\tilde q}^4}
\left[ \frac{4\pi}{\left(16\pi^2 \right)^3}  \hat f_{4PS} \right] \ ,
\eea 
where $\hat f_{4PS}$ is a dimensionless four-body phase-space function. 
Compared to the $\stR$ NLSP decay width (Eq.~\ref{stRGammaf.EQ}),
\beq
c\tau_{\sGl} = c\tau_{\stR} 10^4 \left(\frac{M_\stR M_{\tilde q}}{M_\sH M_\sGl} \right)^4 
\left(\frac{\hat f_{3PS}}{\hat f_{4PS}} \right) \ .
\eeq
We naively estimate $\tau_{\tilde g} \gtrsim 10^4 \tau_{\stR}$, so that, 
for all practical collider purposes, the gluino is stable ($c\tau\gtrsim 100$~m). 
Such a long lived gluino forms an R-hadron and some of its experimental signatures have been
discussed in~\cite{Kraan:2004tz}. 

As an aside, we comment that if the mass spectrum is such that the gluino is no longer so long lived, 
the fact that the gluino is a Majorana particle can be used in order to identify gluino production and decay. Such processes can result 
in decays into a pair of same-sign leptons, and, if kinematically accessible, the case of decays 
into same-sign tops can be easily distinguished from background~\cite{Allanach:2006fy}.

%%%%%%%%%%%%%%%%%%%%%%%%%%%%%%%
\subsection{Gaugino}
Charginos can be pair-produced via off-shell $\gamma$ and $Z$-boson exchange,
and the NLSP chargino decays into the LSP via $\schi^+ \rightarrow \ell^+ \sNR$. This leads
to the signature $p p(\bar p) \rightarrow \ell^+ \ell^- + \mEt$. 
The production cross-section is suppressed relative to that of strongly interacting 
SUSY particle pair production by 
$(g/g_s)^4 \sim 10^{-2}$, and therefore probably too small to be probed at the Tevatron. The LHC, on
the other hand, will have the ability to produce weakly interacting states in significant numbers.
The chargino lifetime can be easily estimated as
\beq
\Gamma_\sH \sim Y_N^2 M_\sH \left[\frac{4\pi}{16\pi^2}\hat{f}_{2PS}\right] \ , 
\eeq
where $\hat f_{2PS}$ is a dimensionless two-body phase-space function.
Comparing this with the stop NLSP decay width (Eq.~\ref{stRGammaf.EQ}), we obtain
\beq
c\tau_{\sH} = c\tau_{\stR} \left(\frac{M_\stR}{M_\sH} \right)^5 
\left(\frac{1}{16\pi^2}\frac{\hat f_{3PS}}{\hat f_{2PS}} \right) \ .
\eeq
For $M_\sH = 1.1 M_\stR$, $\hat{f}_{2PS} \sim O(1)$, and our numerical estimate
$\hat{f}_{3PS} \approx 0.05$, we get $c\tau_\sH \sim c\tau_\stR \times 10^{-4}$. 
Chargino decay is, as far as hadron collider experiments are concerned, prompt. 
The main background is expected to be $W$-boson pair production.  

Neutralinos can be pair-produced via an off-shell squark, and an NLSP neutralino
decays, most of the time, invisibly:  $\schi^0 \rightarrow \nu \sNR$.
A gluon jet or photon can be radiated from the initial state leading to the signature 
$p p (\bar p) \rightarrow j \mEt \ {\rm or}\ \gamma \mEt$. 
Due to the electroweak production cross-section, it 
is unlikely that such processes are accessible at the Tevatron. 
At the LHC, on the other hand, 
one may run into a large number of these events. Whether or not these can be extracted from the
various backgrounds requires a dedicated study, beyond the ambitions of this brief section.

%%%%%%%%%%%%%%%%%%%%%%%%%%%%%%%
\subsection{Slepton}
Sleptons are pair produced by the exchange of virtual $\gamma$ and  $Z$-boson exchange, followed by the 
decay $\tilde\ell \rightarrow \ell \nu \sNR$. This leads to the signature
$p p (\bar p) \rightarrow \ell^+ \ell^- + \mEt$. Since this is a three-body decay, like in
the case of the stop NLSP, we expect displaced vertices or very long-lived sleptons. 
The only observable
decay product of the NLSP slepton is the charged lepton, so the displaced vertex is characterized by a lepton track that 
does not point back to the primary vertex. The dominant background for this channel is $W$-boson pair production.
Note that this case is similar to some manifestations of gauge-mediated SUSY breaking with the
gravitino as the LSP and the slepton (usually the scalar tau), as the NLSP~\cite{GMSBRef}.

%%%%%%%%%%%%%%%%%%%%%%%%%%%%%%%
\subsection{Co-LSP right-handed sneutrino}
Given that there are at least two generations of right-handed sneutrinos ($\sNR^{(i)}$),
we explore the possibility of observable consequences of these ``co-LSP's.''  
Such a possibility was already raised during studies of  the lightest left-handed sneutrino
as the LSP~\cite{left_snu_lsp}. 
It is possible, in all the decays considered so far, that a heavier co-LSP,
say $\sNR^{(2)}$, is produced.  This state later decays to the ``real'' LSP,
say $\sNR^{(1)}$. The relevant (observable) decay channels are 
$\sNR^{(2)} \rightarrow \sNR^{(1)} \ell^+ \ell^-$ and
the one-loop decay $\sNR^{(2)} \rightarrow \sNR^{(1)} \gamma$. 
An estimate of these decay widths is 
\bea
\Gamma_{\sNR^{(2)}} &\sim & Y_N^4 \frac{M_{\sNR^{(2)}}^3}{M_{\tilde\chi}^2} 
 \left[ \frac{4\pi}{(16\pi^2)^2} \hat{f}_{3PS} \right] \ , \\
	&=& 10^{-26} \left(\frac{M_{\sNR^{(2)}}}{100~{\rm GeV}}\right) \left(\frac{M_{\sNR^{(2)}}}{M_{\tilde\chi}}\right)^2 \hat{f}_{3PS} ~ {\rm GeV} \ , \nonumber
\eea
where $\hat{f}_{3PS}$ is a dimensionless phase space factor. Note that this decay width is proportional to $Y_N$ -- tiny! -- to the fourth power. This leads to a lifetime well above $1~$s, so that $\sNR^{(2)}$ is long-lived enough to exit the  detector unseen.

A concern for such long lived particles is whether they disrupt
the successful predictions of big-bang nucleosynthesis. 
Note that, if the mass difference between the
co-LSP states is relatively small (as one would naively expect), $\hat{f}_{3PS} \ll 1$, and the lifetime 
may be orders of magnitude longer than the naive estimate above.
A more more detailed study, beyond the ambitions of this section, is required in order to determine 
the impact of these decaying co-LSPs in the early universe.

%%%%%%%%%%%%%%%%%%%%%%%%%%%%%%%%%%%%%%%%%%%%%%%%%%%%%%%%%%%%%%%%%%%%%%%%%%%%%%%%%%%%%%%
\section{Conclusions}
\label{CONCL.SEC}
Right-handed sneutrinos ($\sNR$) are present in a supersymmetric theory that includes
right-handed neutrinos. If the right-handed neutrino Majorana mass and the SUSY breaking scale are identified 
with the electroweak scale, our understanding of light neutrino masses requires the neutrino Yukawa coupling to be $Y_N \approx 10^{-6}$.
Depending on the details of the SUSY breaking mechanism, it is possible that a predominantly
right-handed sneutrino is the LSP. We detailed such a theory in \cite{Gopalakrishna:2006kr}, 
and explored the cosmological implications of a right-handed 
sneutrino LSP. In this paper we study some hadron collider (Tevatron and LHC) signatures
of such a $\sNR$ LSP. 

We show  that if such a $\sNR$ is the lightest supersymmetric particle (LSP), the collider signatures are 
very interesting, owing to the fact that the $\sNR$ interacts only through the
tiny $Y_N$. If $R$-parity is conserved, all heavier SUSY particles 
 eventually have to cascade decay  to the $\sNR$ through the interaction parameterized by
  $Y_N$. Among other potentially observable effects (leptons, missing energy), we find that the 
  next-to lightest supersymmetric particle (NLSP) is
potentially (very) long-lived. If a mostly right-handed sneutrino is the LSP, one generically 
expects displaced vertices due to the relatively long NLSP lifetime, or heavy, 
collider-stable hadronic or weak states. Since the NLSP decay to the sneutrino LSP proceeds through
a Yukawa coupling, we expect non-universal rates in the $e$, $\mu$ and $\tau$ lepton channels.
Even if the right-handed sneutrinos constituted all of the dark-matter observed, given its
tiny interaction cross-section, the event rate in direct-detection experiments 
will be well below the sensitivity of planned experiments. These aspects help in distinguishing
this scenario from other SUSY models in interpreting any deviations from SM predictions.
Of course, in order to decide whether the scenario discussed here is realized in nature, 
it would be imperative to establish that low-energy SUSY is indeed realized in nature. 
In order to achieve this, it is important to observe a handful of superpartners, and establish 
that their spins and interactions agree with the predictions of a supersymmetric version of 
the standard model. A detailed discussion of this is beyond the ambitions of this paper, but we 
refer readers to~\cite{SUSYestab} for a list of very recent discussions of this issue.

In order to illustrate these aspects, we consider, in Sec.~\ref{stRdec.SEC}, the scalar top
3-body decay $\stR \rightarrow b \ell^+ \sNR$ via an off-shell $\tilde H$.
We calculate the decay matrix element showing explicitly the dependence on $\theta_{b\ell}$,
the angle between the 3-momenta of the final-state $b$-quark and charged lepton.  
We show that for $Y_N \sim 10^{-6}$, $c\tau$ for the scalar top varies from millimeters to meters. 

We performed a simulation of stop pair production and decay at the Tevatron and the LHC, using the Monte Carlo program Pythia.
Pythia by default treats the 3-body final state according only to phase-space, so
we modified the program to correctly incorporate the decay matrix element. 
The signature of this signal is 
$p p (\bar p) \rightarrow \stR \stR^* \rightarrow b \ell^+ \bar{b} \ell^- + \mEt$.
The dominant physics background is top quark pair production 
$p p (\bar{p}) \rightarrow t \bar t \rightarrow b W^+ \bar{b} W^- \rightarrow 
b \ell^+ \bar{b} \ell^- + \mEt$, where the missing energy  is due to final state neutrinos.

The transverse displaced vertex after folding in the boost of the stop is shown in 
Fig.~\ref{dTvx_1.FIG}, and is easily discernible at the Tevatron and the LHC.
This can be used to very effectively suppress the background. If the parameters are
such that a displaced vertex cannot be resolved, we will have to rely on the
distributions and event rates in order to show an excess above background.
We see that the $p_T$ distributions of the $b$-quark and the charged lepton from a 
$225$~GeV stop and the top background are quite similar, and no clear set of $p_T$ cuts can be applied
to separate them. The $\cos\theta_{b\ell}$ distribution is, however, sufficiently different 
for signal and background, as can be seen in Fig.~\ref{cthbE_1.FIG}.

We summarize the Tevatron and LHC reach in Tables~\ref{TevReach.TAB} and~\ref{LHCReach.TAB}. 
From this physics level study, we estimate that the Tevatron can probe stop masses up to about 
$300~$GeV with 10~${\rm fb}^{-1}$, while the LHC is sensitive to about $650~$GeV  
with 500~${\rm fb}^{-1}$ of integrated luminosity. 
This reach was obtained after applying only the level~1 cuts shown in Table~\ref{level1.TAB}. 
A more sophisticated analysis  is expected  to extend the estimates obtained here. Needless to say,
a full  detector-level simulation is necessary in order to realistically assess the capabilities 
of the Tevatron and the LHC. We would like to point out that the results presented in Tables~\ref{TevReach.TAB} and~\ref{LHCReach.TAB} should also apply to other SUSY scenarios in which the scalar top decays predominantly to a $b$-jet, a charged lepton, and missing transverse energy, regardless of whether the stop decay occurs promptly or leaves behind a displaced vertex. 

Of course, the nature of the NLSP depends on the details of supersymmetry breaking. 
We offered some remarks in Sec.~\ref{OthSusy.SEC} on what channels are promising 
for several potential NLSP candidates. 
We leave a comprehensive analysis of many possible cascade decay
chains for future work. 

\vspace*{5mm}
\noindent
{\bf Acknowledgments}

\noindent 
We  thank the organizers and participants of the LC workshop Snowmass 2005, where 
this work was initiated. 
We thank Steve Mrenna for discussions, help with Pythia and comments on the manuscript;
Peter Skands for discussions and help with Pythia;
Marcela Carena, Bob McElrath and Tilman Plehn for discussions; 
Jon Hays, Victoria Martin, Jason Nielsen, and Reinhard Schwienhorst for discussions
on experimental issues. 
The work of AdG and SG is sponsored in part by the US Department of Energy Contract 
DE-FG02-91ER40684, while WP is supported by a 
MEC Ramon y Cajal contract and by the Spanish grant FPA2005-01269.

%%%%%%%%%%%%%%%%%%%%%%%%%%%%%%%%%%%%%%%%%%%%%%%%%%%%%%%%%%%%%%%%%%%%%%%%%%%%%%%%%%%%%%%%
%%%%%%%%%%%%%%%%%%%%%%%%%%%%%%%%%%%%%%%%%%%%%%%%%%%%%%%%%%%%%%%%%%%%%%%%%%%%%%%%%%%%%%%%
\appendix

%%%%%%%%%%%%%%%%%%%%%%%%%%%%%%%%%%%%%%%%%%%%%%%%%%%%%%%%%%%%%%%%%%%%%%%%%%%%%%%%%%%%%%%
\section{The Model}
\label{Theory.SEC}
To the field content of the MSSM, we add (for each generation) a
right-handed neutrino superfield $\widehat N = (\tilde{N}_R, N, F_N)$. Written
as left-chiral fields, the superfields are: $\widehat Q$, $\widehat U^c$, 
$\widehat D^c$, $\widehat L$, $\widehat E^c$, $\widehat N^c$. 
As usual, the MSSM Higgs doublet superfields are $\widehat H_u$
and $\widehat H_d$. Here we repeat the main aspects of the model, details of 
which can be found in \cite{Gopalakrishna:2006kr}.

The superpotential is
\beq
{\cal W} = \widehat U^c Y_U \widehat Q\cdot \widehat H_u
         - \widehat D^c Y_D \widehat Q \cdot \widehat H_d
         + \widehat N^c Y_N \widehat L\cdot \widehat H_u
         - \widehat E^c Y_E \widehat L\cdot \widehat H_d 
		+ \widehat N^c \frac{M_N}{2} \widehat N^c
		+ \mu \widehat H_u \cdot \widehat H_d  \ , 
\label{WSupPot.EQ}
\eeq
where $A \cdot B$ denotes the antisymmetric product of the fields $A$
and $B$, $Y$ are the Yukawa couplings that are $3\times 3$ matrices in
generation space, and, $M_N$ breaks lepton number. 

After electroweak symmetry breaking (when the Higgs scalars get vacuum expectation values $v_u$ and $v_d$), the lowest neutrino mass eigenvalue is given by the standard seesaw relation
\beq
m_{\nu} = \frac{v_u^2 Y_N^2}{M_N} \ ,
\label{mnuss.EQ}
\eeq
where we assume $v_u Y_N \ll M_N$. Neutrino oscillation experiments indicate that 
$m_\nu \sim 0.1~$eV, and if $M_N \sim v$, Eq.~(\ref{mnuss.EQ}) implies that 
$Y_N \sim 10^{-6}$.  

The soft SUSY breaking Lagrangian is given by 
\bea
{\cal L}_{SUSY Br} = &-& \tilde{q}_L^\dagger m^2_{q} \tilde{q}_L 
	- \tilde{u}_R^\dagger m^2_u \tilde{u}_R	- \tilde{d}_R^\dagger m^2_d \tilde{d}_R 
	+ (- \tilde{u}_R^\dagger A_u \tilde{q}_L\cdot h_u 
          + \tilde{d}_R^\dagger A_d \tilde{q}_L\cdot h_d + h.c.) \nonumber \\
	&-& \tilde{\ell}_L^\dagger m^2_{\ell} \tilde{\ell}_L 
	- \tilde{N}_R^\dagger m^2_N \tilde{N}_R	- \tilde{e}_R^\dagger m^2_e \tilde{e}_R 
	+ (- \tilde{N}_R^\dagger A_N \tilde{\ell}_L\cdot h_u
  + \tilde{e}_R^\dagger A_e \tilde{\ell}_L\cdot h_d + h.c.) \nonumber \\
	&+& \left[ (\tilde\ell\cdot h_u)^T \frac{c_{\ell}}{2} (\tilde\ell\cdot h_u) + \tilde{N}_R^T \frac{b_N M_N}{2} \tilde{N}_R + h.c.\right] \nonumber \\
	&+& (b\mu h_u\cdot h_d + h.c.) \ , 
\label{LSUSYBr.EQ}
\eea
where $c_{\ell}$ and $b_N M_N$ are SUSY-breaking, lepton-number breaking parameters. As usual, $m^2$ are SUSY breaking scalar masses-squared, $A$ are SUSY breaking $A$-terms, and $b$ is the  SUSY breaking Higgs boson B-term.

After electroweak symmetry breaking, the sneutrino mass matrix (generation structure suppressed) is given by
\bea
{\cal M}_{\snu} &=& \frac{1}{2}
\pmatrix{m_{LL}^2 & m_{RL}^{2\, \dagger} & -v_u^2 c_\ell^\dagger & v_u Y_N^\dagger M_N \cr 
m_{RL}^2 & m_{RR}^2 & v_u M_N^T Y_N^* & - (b_N M_N)^\dagger \cr 
- v_u^2 c_\ell & v_u Y_N^T M_N^* & m_{LL}^{2\, *} & m_{RL}^{2\, T} \cr 
v_u M_N^\dagger Y_N & - b_N M_N & m_{RL}^{2\, *} & m_{RR}^{2\, *}} \ ,
\label{snumassc.EQ}
\eea
where $m_{LL}^2 = (m_\ell^2 + v_u^2 Y_N^\dagger Y_N + \Delta_\nu^2)$, 
$m_{RR}^2 = (M_N M_N^* + m_N^2 + v_u^2 Y_N Y_N^\dagger)$, 
$m_{RL}^2 = (-\mu^* v_d Y_N + v_u A_N)$, and 
$\Delta_\nu^2 = (m_Z^2/2)\cos{2\beta}$ is the D-term contribution. 

The $\snuL$-$\sNR$ mixing angle is given by (see \cite{Gopalakrishna:2006kr} for 
details\footnote{In Eqs.~(2.5)~and~(2.7) of \cite{Gopalakrishna:2006kr}, the 
term $c_\ell$ should correctly read $v_u^2 c_\ell$, as shown in Eq.~(\ref{thsnu.EQ})
here.})
\bea
\tan{2\theta_{\snu}} &=& \frac{2\left( m_{RL}^2 \pm v_u M_N^\dagger Y_N \right)}{(m_{LL}^2 \mp v_u^2 c_\ell)- (m_{RR}^2 \mp b_N M_N)} \ . \label{thsnu.EQ}
\eea
If $A_N \propto Y_N$, as is the case in several popular SUSY breaking scenarios, it is easy to see that
 $\sin{\theta_{\snu}} \sim Y_N$. For $Y_N \sim 10^{-6}$ -- the case of interest here --
the mixing angle is  tiny, and, as long as $m_{RR}^2 < m_{LL}^2$, the LSP is
an almost pure $\sNR$.

%%%%%%%%%%%%%%%%%%%%%%%%%%%%%%%%%%%%%%%%%%%%%%%%%%%%%%%%%%
\section{The decay $\tilde t_1 \to b l^+ {\tilde \nu}_1$}
\label{AppDecay.SEC}
Here, we present formulas for the decay width of the 3-body decay of the stop, 
$\tilde t_1 \to b l^+ {\tilde \nu}_1$. 
Although in the main body of the paper we consider the case 
when $\theta_\snu \ll 1$ with ${\tilde \nu}_1 \approx \sNR$ as the LSP, the formulas
in this section are valid in general. 

The matrix element is given by
\begin{eqnarray}
T_{fi} &=& \sum_{i=1,2} \bar{u}_b(p_b) \left( \kt{i1} P_L + \lt{i1} P_R \right)
         \frac{1}{ \pslash{\chip{i}} - m_{\chip{i}}}
       \left( \ksnc{i1} P_R + \lsnc{i1} P_L \right) v_\tau(p_\tau)\ ,
\end{eqnarray}
where 
\begin{eqnarray}
\begin{array}{lcl}
\lt{i1} = -g \cos \theta_{\tilde t} V_{i1} - Y_t \sin \theta_{\tilde t} V_{i2}
 &, & \kt{i1} =  Y_b \cos \theta_{\tilde t} U_{i2} \ , \\
\lsn{i1} =  -g \cos \theta_{\tilde \nu} V_{i1} - Y_\nu \sin \theta_{\tilde \nu} V_{i2}
 &, & \ksn{i1} =  Y_\tau \cos \theta_{\tilde \nu} U_{i2} \ , \\
\end{array}
\end{eqnarray}
in the case of Dirac neutrinos. In the case of Majorana neutrinos,
$\lsn{i1}$ and $\ksn{i1}$ have to be multiplied by $1/\sqrt{2}$ or
$i/\sqrt{2}$ for the cases that $\tilde \nu_1$ is either the scalar
state or the pseudo-scalar state, respectively. 
The total width is given by 
\begin{eqnarray}
\Gamma(\tilde t_1 \to b \tau^+ \tilde \nu_1)
 &=& \frac{1}{16 m_{{\tilde t}_1} (2 \pi)^5}
\int \frac{d^3 \, p_b}{E_b}  \frac{d^3 \, p_\tau}{E_\tau}
\frac{\delta(E_{{\tilde t}_1} - E_b - E_\tau - E_{{\tilde \nu}_1})}
     {E_{{\tilde \nu}_1}} \left| T_{fi} \right|^2 \ ,
\end{eqnarray}
with
\begin{eqnarray}
E_{{\tilde \nu}_1} &=& 
\sqrt{m^2_{{\tilde \nu}_1} + (\vec{p}_{{\tilde t}_1} -\vec{p}_b -\vec{p}_\tau)^2} \ ,
\\
\left| T_{fi} \right|^2 &=&
 \sum_{i=1,2} \frac{Tr_{ii}}
                   {\left((p_{{\tilde t}_1} - p_b)^2 - m^2_{\chip{i}}\right)^2}
 + 2 {\mathrm Re} \left(\frac{Tr_{12}}
                   {\left((p_{{\tilde t}_1} - p_b)^2 - m^2_{\chip{1}}\right)
                    \left((p_{{\tilde t}_1} - p_b)^2 - m^2_{\chip{2}}\right)}
           \right) \ ,
\nonumber \\ \\
Tr_{ii} &=&  2 \left( |\kt{i1}|^2  |\ksn{i1}|^2 
         + |\lt{i1}|^2 |\lsn{i1}|^2  \right)
 \left( 2 (p_{\tilde t_1} - p_b) \cdot p_\tau \,\,\,
          (p_{\tilde t_1} - p_b) \cdot p_b
      - (p_{\tilde t_1} - p_b)^2 \,\,\, p_b\cdot p_\tau \right)
\nonumber \\ &+&   2 \,   m^2_{\chip{i}}  
  \left( |\kt{i1}|^2  |\lsn{i1}|^2 
               + |\lt{i1}|^2  |\ksn{i1}|^2  \right)
   p_{\tau} \cdot p_{b} 
\nonumber \\ &+& 4 \, m_b m_{\chip{i}} {\mathrm Re}(\kt{i1} \ltc{i1} )
  \left( |\lsn{i1}|^2 +  |\ksn{i1}|^2  \right)
      (p_{\tilde t_1} - p_b) \cdot p_\tau
\nonumber \\ &-&  4 \,    m_\tau  m_{\chip{i}} {\mathrm Re}(\ksn{i1} \lsnc{i1} )
  \left( |\lt{i1}|^2 +  |\kt{i1}|^2  \right)
    (p_{\tilde t_1} - p_b) \cdot p_b
\nonumber \\ &-& 4 \,   m_\tau m_b  {\mathrm Re}
  \left( \kt{i1}  \ksnc{i1} \lsn{i1} \ltc{i1} \right)
 \left( (p_{\tilde t_1} - p_b) \cdot (p_{\tilde t_1} - p_b) +  m^2_{\chip{i}}
 \right) \ ,
\label{eq:Tii} \\
Tr_{12} &=& 2 \left( \kt{11}  \ksnc{11}  \ksn{21}  \ktc{21}
         + \lt{11} \lsnc{11} \lsn{21} \ltc{21} \right) \times 
 \nonumber \\ &&
 \left( 2 (p_{\tilde t_1} - p_b) \cdot p_\tau \,\,\,
          (p_{\tilde t_1} - p_b) \cdot p_b
      - (p_{\tilde t_1} - p_b)^2 \,\,\, p_b\cdot p_\tau \right)
\nonumber \\ &+&   2 \,   m_{\chip{1}}  m_{\chip{2}}
  \left( \kt{11} \lsnc{11} \lsn{21} \ktc{21} 
               + \lt{11}  \ksnc{11} \ltc{21} \ksn{21}  \right)
   p_{\tau} \cdot p_{b} 
\nonumber \\ &+& 2 \, m_b m_{\chip{1}}
  \left( \kt{11} \lsnc{11} \lsn{21} \ltc{21}
        + \lt{11} \ksnc{11}  \ksn{21}  \ktc{21}  \right)
      (p_{\tilde t_1} - p_b) \cdot p_\tau
\nonumber \\ &+&   2 \,   m_{\chip{2}} m_b
  \left( \kt{11}  \ksnc{11} \ltc{21} \ksn{21} 
               + \lt{11}  \lsnc{11} \lsn{21} \ktc{21}  \right)
   (p_{\tilde t_1} - p_b) \cdot p_{\tau} 
\nonumber \\ &-&  2 \,    m_\tau  m_{\chip{1}}
 \left( \kt{11}\lsnc{11} \ksn{21}  \ktc{21}
               + \lt{11}  \ksnc{11} \lsn{21} \ltc{21}  \right)
    (p_{\tilde t_1} - p_b) \cdot p_b
\nonumber \\ &-&   2 \,   m_{\chip{2}}  m_\tau
 \left( \kt{11}  \ksnc{11} \lsn{21} \ktc{21} 
               + \lt{11}\lsnc{11}  \ltc{21} \ksn{21} \right)
   (p_{\tilde t_1} - p_b) \cdot p_{b} 
\nonumber \\ &-& 2 \,   m_\tau m_b
  \left( \kt{11}  \ksnc{11} \lsn{21} \ltc{21}
               + \lt{11} \lsnc{11} \ksn{21}  \ktc{21} \right)
  (p_{\tilde t_1} - p_b) \cdot (p_{\tilde t_1} - p_b)
\nonumber \\ &-&   2 \,   m_{\chip{2}}  m_\tau m_b  m_{\chip{1}}
 \left( \kt{11}\lsnc{11}  \ltc{21} \ksn{21}
        + \lt{11}  \ksnc{11} \lsn{21} \ktc{21} \right) \ .
\label{eq:T12}
\end{eqnarray}

%%%%%%%%%%%%%%%%%%%%%%%%%%%%%%%%%%%%%%%%%%%%%%%%%%%%%%%%%%%%%%%%%%%%%%%%


\begin{thebibliography}{99}

%\cite{Minkowski:1977sc}
\bibitem{Minkowski:1977sc}
  P.~Minkowski,
  %``Mu $\to$ E Gamma At A Rate Of One Out Of 1-Billion Muon Decays?,''
  Phys.\ Lett.\ B {\bf 67}, 421  (1977);
  %%CITATION = PHLTA,B67,421;%%
M.~Gell-Mann, P.~Ramond, and R.~Slansky,
``Complex Spinors and Unified Theories'', Proceedings of the
  Workshop, Stony Brook, New York, North-Holland, 1979;
%
T.~Yanagida, (KEK, Tsuhuba), 1979;
%
See also 
  R.~N.~Mohapatra and G.~Senjanovic,
  %``Neutrino Mass And Spontaneous Parity Nonconservation,''
  Phys.\ Rev.\ Lett.\  {\bf 44}, 912 (1980);
  %%CITATION = PRLTA,44,912;%%
%\cite{Schechter:1980gr}
%\bibitem{Schechter:1980gr}
  J.~Schechter and J.~W.~F.~Valle,
  %``Neutrino Masses In SU(2) X U(1) Theories,''
  Phys.\ Rev.\ D {\bf 22}, 2227 (1980).
  %%CITATION = PHRVA,D22,2227;%%

\bibitem{osc_review}
For recent, complete reviews (and many references), see
A.~de Gouv\^ea,
  %``2004 TASI lectures on neutrino physics,''
  [hep-ph/0411274];
  %%CITATION = HEP-PH 0411274;%%
 A.~Strumia and F.~Vissani,
  %``Neutrino masses and mixings and,''
 [hep-ph/0606054].
  %%CITATION = HEP-PH 0606054;%%

%\cite{Blair:2002pg}
\bibitem{Blair:2002pg}
  G.~A.~Blair, W.~Porod and P.~M.~Zerwas,
  %``The reconstruction of supersymmetric theories at high energy scales.
  %((U)),''
  Eur.\ Phys.\ J.\ C {\bf 27}, 263 (2003)
  [hep-ph/0210058];
  %%CITATION = HEP-PH 0210058;%%
 A.~Freitas, W.~Porod and P.~M.~Zerwas,
  %``Determining sneutrino masses and physical implications,''
  Phys.\ Rev.\ D {\bf 72}, 115002 (2005)
  [hep-ph/0509056];
  %%CITATION = HEP-PH 0509056;%%
%
  M.~R.~Buckley and H.~Murayama,
  %``How can we test seesaw experimentally?,''
  [hep-ph/0606088].
  %%CITATION = HEP-PH 0606088;%%

\bibitem{SUSYseesaw}
%\cite{Hisano:1995nq}
%\bibitem{Hisano:1995nq}
FCNC effects have been considered in, for example:
  J.~Hisano, T.~Moroi, K.~Tobe, M.~Yamaguchi and T.~Yanagida,
%   ``Lepton flavor violation in the supersymmetric standard model with seesaw
% induced neutrino masses,''
  Phys.\ Lett.\ B {\bf 357}, 579 (1995)
  [hep-ph/9501407];
  %%CITATION = HEP-PH 9501407;%%
%\cite{Hisano:1995cp}
%\bibitem{Hisano:1995cp}
  J.~Hisano, T.~Moroi, K.~Tobe and M.~Yamaguchi,
%   ``Lepton-Flavor Violation via Right-Handed Neutrino Yukawa Couplings in
% Supersymmetric Standard Model,''
  Phys.\ Rev.\ D {\bf 53}, 2442 (1996)
  [hep-ph/9510309];
  %%CITATION = HEP-PH 9510309;%%
%\cite{Hirsch:1997vz}
%\bibitem{Hirsch:1997vz}
  M.~Hirsch, H.~V.~Klapdor-Kleingrothaus and S.~G.~Kovalenko,
  %``B-L violating masses in softly broken supersymmetry,''
  Phys.\ Lett.\ B {\bf 398}, 311 (1997)
  [hep-ph/9701253];
  %%CITATION = HEP-PH 9701253;%%
%\cite{Hirsch:1998gr}
%\bibitem{Hirsch:1998gr}
  M.~Hirsch, H.~V.~Klapdor-Kleingrothaus, S.~Kolb and S.~G.~Kovalenko,
  % ``Phenomenological implications of 'Majorana' sneutrinos at future
  %accelerators,''
  Phys.\ Rev.\ D {\bf 57}, 2020 (1998);
  %%CITATION = PHRVA,D57,2020;%%
%\cite{Ellis:2002fe}
%\bibitem{Ellis:2002fe}
  J.~R.~Ellis, J.~Hisano, M.~Raidal and Y.~Shimizu,
%   ``A new parametrization of the seesaw mechanism and applications in
% supersymmetric models,''
  Phys.\ Rev.\ D {\bf 66}, 115013 (2002)
  [hep-ph/0206110];
  %\cite{Masiero:2004js}
%\bibitem{Masiero:2004js}
  A.~Masiero, S.~K.~Vempati and O.~Vives,
  %``Massive neutrinos and flavour violation,''
  New J.\ Phys.\  {\bf 6}, 202 (2004)
  [hep-ph/0407325].
  %%CITATION = HEP-PH 0407325;%%
%%CITATION = HEP-PH 0206110;%%
Sneutrino-antisneutrino mixing has been considered in:
%\cite{Grossman:1997is}
%\bibitem{Grossman:1997is}
  Y.~Grossman and H.~E.~Haber,
  %``Sneutrino mixing phenomena,''
  Phys.\ Rev.\ Lett.\  {\bf 78}, 3438 (1997)
  [hep-ph/9702421].
  %%CITATION = HEP-PH 9702421;%%

\bibitem{typeIIrefs}
See the last reference in \cite{Minkowski:1977sc} above;
%\cite{Gelmini:1980re}
%\bibitem{Gelmini:1980re}
  G.~B.~Gelmini and M.~Roncadelli,
  %``Left-Handed Neutrino Mass Scale And Spontaneously Broken Lepton Number,''
  Phys.\ Lett.\ B {\bf 99}, 411 (1981);
  %%CITATION = PHLTA,B99,411;%%
For a discussion of some interesting phenomenological consequences, albeit non-supersymmetric, see for example:
  %\cite{Ma:1998dx}
%\bibitem{Ma:1998dx}
 E.~Ma and U.~Sarkar,
  %``Neutrino masses and leptogenesis with heavy Higgs triplets,''
  Phys.\ Rev.\ Lett.\  {\bf 80}, 5716 (1998) 
  [hep-ph/9802445];
  %%CITATION = HEP-PH 9802445;%%
%\cite{deGouvea:2005jj}
%\bibitem{deGouvea:2005jj}
  A.~de Gouvea and S.~Gopalakrishna,
%   ``Low-energy neutrino Majorana phases and charged-lepton electric dipole
  %moments,''
  Phys.\ Rev.\ D {\bf 72}, 093008 (2005)
  [hep-ph/0508148].
  %%CITATION = HEP-PH 0508148;%%

%\cite{deGouvea:2005er}
\bibitem{deGouvea:2005er}
  A.~de Gouv\^ea,
  %``See-saw energy scale and the LSND anomaly,''
  Phys.\ Rev.\ D {\bf 72}, 033005 (2005)
  [hep-ph/0501039].
  %%CITATION = HEP-PH 0501039;%%

%\cite{Gopalakrishna:2006kr}
\bibitem{Gopalakrishna:2006kr}
  S.~Gopalakrishna, A.~de Gouv\^ea and W.~Porod,
  %``Right-handed sneutrinos as nonthermal dark matter,''
  JCAP {\bf 0605}, 005 (2006)
  [hep-ph/0602027].
  %%CITATION = HEP-PH 0602027;%%

%\cite{Garbrecht:2006az}
\bibitem{Garbrecht:2006az}
  B.~Garbrecht, C.~Pallis and A.~Pilaftsis,
  %``Anatomy of F(D)-term hybrid inflation,''
  [hep-ph/0605264].
  %%CITATION = HEP-PH 0605264;%%

%\cite{Asaka:2005cn}
\bibitem{Asaka:2005cn}
  T.~Asaka, K.~Ishiwata and T.~Moroi,
  %``Right-Handed Sneutrino as Cold Dark Matter,''
  [hep-ph/0512118].
  %%CITATION = HEP-PH 0512118;%%

\bibitem{SnuLRefs}
%\cite{Hagelin:1984wv}
%\bibitem{Hagelin:1984wv}
  J.~S.~Hagelin, G.~L.~Kane and S.~Raby,
  %``Perhaps Scalar Neutrinos Are The Lightest Supersymmetric Partners,''
  Nucl.\ Phys.\ B {\bf 241}, 638 (1984);
  %%CITATION = NUPHA,B241,638;%%
%\cite{Hall:1997ah}
%\bibitem{Hall:1997ah}
  L.~J.~Hall, T.~Moroi and H.~Murayama,
  %``Sneutrino cold dark matter with lepton-number violation,''
  Phys.\ Lett.\ B {\bf 424}, 305 (1998)
  [hep-ph/9712515];
  %%CITATION = HEP-PH 9712515;%%
%\cite{Kolb:1999wx}
%\bibitem{Kolb:1999wx}
  S.~Kolb, M.~Hirsch, H.~V.~Klapdor-Kleingrothaus and O.~Panella,
  %``Collider signatures of sneutrino cold dark matter,''
  Phys.\ Lett.\ B {\bf 478}, 262 (2000)
  [hep-ph/9910542];
  %%CITATION = HEP-PH 9910542;%%
%\cite{Arkani-Hamed:2000bq}
%\bibitem{Arkani-Hamed:2000bq}
  N.~Arkani-Hamed, L.~J.~Hall, H.~Murayama, D.~R.~Smith and N.~Weiner,
  %``Small neutrino masses from supersymmetry breaking,''
  Phys.\ Rev.\ D {\bf 64}, 115011 (2001)
  [hep-ph/0006312];
  %%CITATION = HEP-PH 0006312;%%
%\cite{Chou:2000cy}
%\bibitem{Chou:2000cy}
  C.~L.~Chou, H.~L.~Lai and C.~P.~Yuan,
  %``New production mechanism of neutral Higgs bosons with right scalar tau
  %neutrino as the LSP,''
  Phys.\ Lett.\ B {\bf 489}, 163 (2000)
  [hep-ph/0006313].
  %%CITATION = HEP-PH 0006313;%%

%\cite{Chou:1999zb}
\bibitem{Chou:1999zb}
  C.~L.~Chou and M.~E.~Peskin,
  %``Scalar top quark as the next-to-lightest supersymmetric particle,''
  Phys.\ Rev.\ D {\bf 61}, 055004 (2000)
  [hep-ph/9909536];
  %%CITATION = HEP-PH 9909536;%%

\bibitem{GMSBRef}
See for example,
 %\cite{Feng:1997zr}
%\bibitem{Feng:1997zr}
  J.~L.~Feng and T.~Moroi,
  %``Tevatron signatures of long-lived charged sleptons in gauge-mediated
  %supersymmetry breaking models,''
  Phys.\ Rev.\ D {\bf 58}, 035001 (1998)
  [hep-ph/9712499];
  %%CITATION = HEP-PH 9712499;%%
%\cite{Giudice:1998bp}
%\bibitem{Giudice:1998bp}
  G.~F.~Giudice and R.~Rattazzi,
  %``Theories with gauge-mediated supersymmetry breaking,''
  Phys.\ Rept.\  {\bf 322}, 419 (1999)
  [hep-ph/9801271].
  %%CITATION = HEP-PH 9801271;%%

%\cite{Strassler:2006im}
\bibitem{Strassler:2006im}
  M.~J.~Strassler and K.~M.~Zurek,
  %``Echoes of a hidden valley at hadron colliders,''
  [hep-ph/0604261].
  %%CITATION = HEP-PH 0604261;%%

\bibitem{LtStRef}
See for example,
%\cite{Ellis:1983ed}
%\bibitem{Ellis:1983ed}
  J.~R.~Ellis and S.~Rudaz,
  %``Search For Supersymmetry In Toponium Decays,''
  Phys.\ Lett.\ B {\bf 128}, 248 (1983);
  %%CITATION = PHLTA,B128,248;%%
%\cite{Barger:1993gh}
%\bibitem{Barger:1993gh}
  V.~D.~Barger, M.~S.~Berger and P.~Ohmann,
  %``The Supersymmetric particle spectrum,''
  Phys.\ Rev.\ D {\bf 49}, 4908 (1994)
  [hep-ph/9311269];
  %%CITATION = HEP-PH 9311269;%%
%\cite{Djouadi:1996pj}
%\bibitem{Djouadi:1996pj}
  A.~Djouadi, J.~Kalinowski, P.~Ohmann and P.~M.~Zerwas,
  %``Heavy SUSY Higgs bosons at e+ e- linear colliders,''
  Z.\ Phys.\ C {\bf 74}, 93 (1997)
  [hep-ph/9605339].
  %%CITATION = HEP-PH 9605339;%%

\bibitem{EWBaryoRefs}
See for example, 
%\cite{Carena:1996wj}
%\bibitem{Carena:1996wj}
  M.~Carena, M.~Quiros and C.~E.~M.~Wagner,
  %``Opening the Window for Electroweak Baryogenesis,''
  Phys.\ Lett.\ B {\bf 380}, 81 (1996)
  [hep-ph/9603420];
  %%CITATION = HEP-PH 9603420;%%
%\cite{Quiros:2000wk}
%\bibitem{Quiros:2000wk}
  M.~Quiros,
  %``Electroweak baryogenesis and the Higgs and stop masses,''
  Nucl.\ Phys.\ Proc.\ Suppl.\  {\bf 101}, 401 (2001)
  [hep-ph/0101230];
  %%CITATION = HEP-PH 0101230;%%
%\cite{Carena:2002ss}
%\bibitem{Carena:2002ss}
  M.~Carena, M.~Quiros, M.~Seco and C.~E.~M.~Wagner,
  %``Improved results in supersymmetric electroweak baryogenesis,''
  Nucl.\ Phys.\ B {\bf 650}, 24 (2003)
  [hep-ph/0208043].
  %%CITATION = HEP-PH 0208043;%%

\bibitem{StopPDRef}
%\cite{Porod:1998yp}
%\bibitem{Porod:1998yp}
W.~Porod and T.~W\"ohrmann,
  %``Higher order top squark decays,''
  Phys.\ Rev.\ D {\bf 55} (1997) 2907
  [Erratum-ibid.\ D {\bf 67} (2003) 059902]
  [hep-ph/9608472];
  W.~Porod,
  %``More on higher order decays of the lighter top squark,''
  Phys.\ Rev.\ D {\bf 59}, 095009 (1999)
  [hep-ph/9812230];
  %%CITATION = HEP-PH 9812230;%%
%\cite{Boehm:1999tr}
%\bibitem{Boehm:1999tr}
  C.~Boehm, A.~Djouadi and Y.~Mambrini,
  %``Decays of the lightest top squark,''
  Phys.\ Rev.\ D {\bf 61} (2000) 095006
  [hep-ph/9907428];
  %%CITATION = HEP-PH 9907428;%%
%\cite{Demina:1999ty}
%\bibitem{Demina:1999ty}
  R.~Demina, J.~D.~Lykken, K.~T.~Matchev and A.~Nomerotski,
  %``Stop and sbottom searches in Run II of the Fermilab Tevatron,''
  Phys.\ Rev.\ D {\bf 62}, 035011 (2000)
  [hep-ph/9910275];
  %%CITATION = HEP-PH 9910275;%%
%\cite{Djouadi:2001dx}
%\bibitem{Djouadi:2001dx}
  A.~Djouadi, M.~Guchait and Y.~Mambrini,
  %``Scalar top quarks at the Run II of the Tevatron in the high tan(beta)
  %regime,''
  Phys.\ Rev.\ D {\bf 64} (2001) 095014
  [hep-ph/0105108];
  %%CITATION = HEP-PH 0105108;%%
%\cite{Das:2001kd}
%\bibitem{Das:2001kd}
  S.~P.~Das, A.~Datta and M.~Guchait,
  %``Four body decay of the stop squark at the upgraded Tevatron,''
  Phys.\ Rev.\ D {\bf 65} (2002) 095006
  [hep-ph/0112182];
  %%CITATION = HEP-PH 0112182;%%
%\cite{Carena:2002wz}
%\bibitem{Carena:2002wz}
  M.~Carena, D.~Choudhury, R.~A.~Diaz, H.~E.~Logan and C.~E.~M.~Wagner,
  %``Top-squark searches at the Tevatron in models of low-energy  supersymmetry
  %breaking,''
  Phys.\ Rev.\ D {\bf 66}, 115010 (2002)
  [hep-ph/0206167].
  %%CITATION = HEP-PH 0206167;%%

%\cite{Allanach:2006fy}
\bibitem{Allanach:2006fy}
  B.~C.~Allanach {\it et al.},
  %``Les Houches 'Physics at TeV colliders 2005' Beyond the standard model
  %working group: Summary report,''
  [hep-ph/0602198].
  %%CITATION = HEP-PH 0602198;%%

\bibitem{PythiaRef}
T.~Sjostrand, P.~Eden, C.~Friberg, L.~Lonnblad, G.~Miu, S.~Mrenna and E.~Norrbin, 
Computer Physics Commun. 135 (2001) 238;
% \cite{Sjostrand:2003wg}
%\bibitem{Sjostrand:2003wg}
  T.~Sjostrand, L.~Lonnblad, S.~Mrenna and P.~Skands,
  %``PYTHIA 6.3: Physics and manual,''
  [hep-ph/0308153].
  %%CITATION = HEP-PH 0308153;%%


%\cite{Kraan:2004tz}
\bibitem{Kraan:2004tz}
  See, for example, A.~C.~Kraan,
  %``Interactions of heavy stable hadronizing particles,''
  Eur.\ Phys.\ J.\ C {\bf 37}, 91 (2004)
  [hep-ex/0404001];
  %%CITATION = HEP-EX 0404001;%%
W.~Kilian, T.~Plehn, P.~Richardson and E.~Schmidt,
  %``Split supersymmetry at colliders,''
  Eur.\ Phys.\ J.\ C {\bf 39}, 229 (2005)
  [hep-ph/0408088];
  %%CITATION = HEP-PH 0408088;%%
J.~L.~Hewett, B.~Lillie, M.~Masip and T.~G.~Rizzo,
  %``Signatures of long-lived gluinos in split supersymmetry,''
  JHEP {\bf 0409}, 070 (2004)
  [hep-ph/0408248];
  %%CITATION = HEP-PH 0408248;%%
A.~C.~Kraan, J.~B.~Hansen and P.~Nevski,
  %``Discovery potential of R-hadrons with the ATLAS detector,''
  hep-ex/0511014.
  %%CITATION = HEP-EX 0511014;%%

\bibitem{gluinonium}
  For a recent discussion (in a rather different context), see
  K.~Cheung and W.~Y.~Keung,
  %``Split supersymmetry, stable gluino, and gluinonium,''
  Phys.\ Rev.\ D {\bf 71}, 015015 (2005)
  [hep-ph/0408335].
  %%CITATION = HEP-PH 0408335;%%


\bibitem{super-onium} 
C.~R.~Nappi,
  %``Spin 0 Quarks In E+ E- Annihilation,''
  Phys.\ Rev.\ D {\bf 25}, 84 (1982).
  %%CITATION = PHRVA,D25,84;%%
  For collider studies see, for example, 
  P.~Moxhay and R.~W.~Robinett,
  %``Searching For Scalar Quarkonium At Proton - Anti-Proton Colliders,''
  Phys.\ Rev.\ D {\bf 32}, 300 (1985);
  %%CITATION = PHRVA,D32,300;%%
  M.~J.~Herrero, A.~Mendez and T.~G.~Rizzo,
  %``Production Of Heavy Squarkonium At High-Energy P P Colliders,''
  Phys.\ Lett.\ B {\bf 200}, 205 (1988);
  %%CITATION = PHLTA,B200,205;%%
  T.~G.~Rizzo,
  %``Squark And Squarkonium Production By Gauge Boson Fusion At Tev E+ E-
  %Colliders,''
  Phys.\ Rev.\ D {\bf 40}, 2803 (1989);
  %%CITATION = PHRVA,D40,2803;%%
  %\cite{Barger:1988sp}
  %\bibitem{Barger:1988sp}
  V.~D.~Barger and W.~Y.~Keung,
  %``Stoponium Decays To Higgs Bosons,''
  Phys.\ Lett.\ B {\bf 211}, 355 (1988);
  %%CITATION = PHLTA,B211,355;%%
  %\cite{Drees:1993yr}
  %\bibitem{Drees:1993yr}
  M.~Drees and M.~M.~Nojiri,
  %``A new signal for scalar top bound state production,''
  Phys.\ Rev.\ Lett.\  {\bf 72}, 2324 (1994)
  [hep-ph/9310209];
  %%CITATION = HEP-PH 9310209;%%
  %\cite{Drees:1993uw}
  %\bibitem{Drees:1993uw}
  M.~Drees and M.~M.~Nojiri,
  %``Production and decay of scalar stoponium bound states,''
  Phys.\ Rev.\ D {\bf 49}, 4595 (1994)
  [hep-ph/9312213].
  %%CITATION = HEP-PH 9312213;%%
  
  

%\cite{Barger:1988jj}
\bibitem{Barger:1988jj}
  V.~D.~Barger, J.~Ohnemus and R.~J.~N.~Phillips,
  %``Spin Correlation Effects In The Hadroproduction And Decay Of Very Heavy Top
  %Quark Pairs,''
  Int.\ J.\ Mod.\ Phys.\ A {\bf 4}, 617 (1989).
  %%CITATION = IMPAE,A4,617;%%

\bibitem{SpinCorrRefs}
%\cite{Kane:1991bg}
%\bibitem{Kane:1991bg}
  G.~L.~Kane, G.~A.~Ladinsky and C.~P.~Yuan,
  %``Using the top quark for testing standard model polarization and CP
  %predictions,''
  Phys.\ Rev.\ D {\bf 45}, 124 (1992);
  %%CITATION = PHRVA,D45,124;%%
%\cite{Stelzer:1995gc}
%\bibitem{Stelzer:1995gc}
  T.~Stelzer and S.~Willenbrock,
  %``Spin Correlation in Top-Quark Production at Hadron Colliders,''
  Phys.\ Lett.\ B {\bf 374}, 169 (1996)
  [hep-ph/9512292];
  %%CITATION = HEP-PH 9512292;%%
%\cite{Chang:1995ay}
%\bibitem{Chang:1995ay}
  D.~Chang, S.~C.~Lee and A.~Sumarokov,
  %``On the Observation of Top Spin Correlation Effect at Tevatron,''
  Phys.\ Rev.\ Lett.\  {\bf 77}, 1218 (1996)
  [hep-ph/9512417];
  %%CITATION = HEP-PH 9512417;%%
%\cite{Bernreuther:2004jv}
%\bibitem{Bernreuther:2004jv}
  W.~Bernreuther, A.~Brandenburg, Z.~G.~Si and P.~Uwer,
  %``Top quark pair production and decay at hadron colliders,''
  Nucl.\ Phys.\ B {\bf 690}, 81 (2004)
  [hep-ph/0403035];
  %%CITATION = HEP-PH 0403035;%%
  For an ATLAS study, see for example, K.~Smolek and V.~Simak, ATL-PHYS-2003-012 (2003).
	
%\cite{Beenakker:1997ut}
\bibitem{Beenakker:1997ut}
See for example 
  W.~Beenakker, M.~Kramer, T.~Plehn, M.~Spira and P.~M.~Zerwas,
  %``Stop production at hadron colliders,''
  Nucl.\ Phys.\ B {\bf 515}, 3 (1998)
  [hep-ph/9710451],
  %%CITATION = HEP-PH 9710451;%%
and references therein.

%\cite{Hesselbach:2000qw}
\bibitem{Hesselbach:2000qw}
  S.~Hesselbach, F.~Franke and H.~Fraas,
  %``Displaced vertices in extended supersymmetric models,''
  Phys.\ Lett.\ B {\bf 492}, 140 (2000)
  [hep-ph/0007310].
  %%CITATION = HEP-PH 0007310;%%

%\cite{deCampos:2005ri}
\bibitem{deCampos:2005ri}
  W.~Porod, M.~Hirsch, J.~Romao and J.~W.~F.~Valle,
  %``Testing neutrino mixing at future collider experiments,''
  Phys.\ Rev.\ D {\bf 63}, 115004 (2001)
  [hep-ph/0011248];
  %%CITATION = HEP-PH 0011248;%%
%
  F.~de Campos, O.~J.~P.~Eboli, M.~B.~Magro, W.~Porod, D.~Restrepo and J.~W.~F.~Valle,
  %``Probing neutrino mass with displaced vertices at the Tevatron,''
  Phys.\ Rev.\ D {\bf 71}, 075001 (2005)
  [hep-ph/0501153];
  %%CITATION = HEP-PH 0501153;%%
%
  M.~Hirsch, W.~Porod and D.~Restrepo,
  %``Collider signals of gravitino dark matter in bilinearly broken  R-parity,''
  JHEP {\bf 0503}, 062 (2005)
  [hep-ph/0503059].
  %%CITATION = HEP-PH 0503059;%%

\bibitem{DispVtxRefs}
%\cite{Ambrosanio:1997bq}
%\bibitem{Ambrosanio:1997bq}
  S.~Ambrosanio, G.~D.~Kribs and S.~P.~Martin,
  % ``Three-body decays of selectrons and smuons in low-energy supersymmetry
  %breaking models,''
  Nucl.\ Phys.\ B {\bf 516}, 55 (1998)
  [hep-ph/9710217];
  %%CITATION = HEP-PH 9710217;%%
%\cite{Ellwanger:1998vi}
%\bibitem{Ellwanger:1998vi}
  U.~Ellwanger and C.~Hugonie,
  %``Topologies Of The (M+1)Ssm With A Singlino Lsp At Lep2,''
  Eur.\ Phys.\ J.\ C {\bf 13}, 681 (2000)
  [hep-ph/9812427];
  %%CITATION = HEP-PH 9812427;%%
%\cite{Feng:1999fu}
%\bibitem{Feng:1999fu}
  J.~L.~Feng, T.~Moroi, L.~Randall, M.~Strassler and S.~f.~Su,
  %``Discovering supersymmetry at the Tevatron in Wino LSP scenarios,''
  Phys.\ Rev.\ Lett.\  {\bf 83}, 1731 (1999)
  [hep-ph/9904250];
  %%CITATION = HEP-PH 9904250;%%
%\cite{Gherghetta:1999sw}
%\bibitem{Gherghetta:1999sw}
  T.~Gherghetta, G.~F.~Giudice and J.~D.~Wells,
  % ``Phenomenological Consequences Of Supersymmetry With Anomaly-Induced
  %Masses,''
  Nucl.\ Phys.\ B {\bf 559}, 27 (1999)
  [hep-ph/9904378];
  %%CITATION = HEP-PH 9904378;%%
%\cite{Martin:2000eq}
%\bibitem{Martin:2000eq}
  S.~P.~Martin,
  % ``Collider signals from slow decays in supersymmetric models with an
  %intermediate-scale solution to the mu problem,''
  Phys.\ Rev.\ D {\bf 62}, 095008 (2000)
  [hep-ph/0005116];
  %%CITATION = HEP-PH 0005116;%%
%\cite{Hesselbach:2000qw}
%\bibitem{Hesselbach:2000qw}
  S.~Hesselbach, F.~Franke and H.~Fraas,
  %``Displaced vertices in extended supersymmetric models,''
  Phys.\ Lett.\ B {\bf 492}, 140 (2000)
  [hep-ph/0007310].
  %%CITATION = HEP-PH 0007310;%%

%\cite{Kidonakis:2004hr}
\bibitem{Kidonakis:2004hr}
  N.~Kidonakis and R.~Vogt,
  %``Theoretical status of the top quark cross section,''
  Int.\ J.\ Mod.\ Phys.\ A {\bf 20}, 3171 (2005)
  [hep-ph/0410367].
  %%CITATION = HEP-PH 0410367;%%

%\cite{Abel:2000vs}
\bibitem{Abel:2000vs}
  S.~Abel {\it et al.}  [SUGRA Working Group Collaboration],
  %``Report of the SUGRA working group for run II of the Tevatron,''
  hep-ph/0003154.
  %%CITATION = HEP-PH 0003154;%%

\bibitem{top_exptRef}
%\cite{Acosta:2004uw}
%\bibitem{Acosta:2004uw}
  D.~Acosta {\it et al.}  [CDF Collaboration],
  %``Measurement of the t anti-t production cross section in p anti-p
  %collisions at s**(1/2) = 1.96-TeV using dilepton events,''
  Phys.\ Rev.\ Lett.\  {\bf 93}, 142001 (2004)
  [hep-ex/0404036];
  %%CITATION = HEP-EX 0404036;%%
%\cite{Abazov:2005yt}
%\bibitem{Abazov:2005yt}
  V.~M.~Abazov {\it et al.}  [D0 Collaboration],
  %``Measurement of the t anti-t production cross section in p anti-p
  %collisions at s**(1/2) = 1.96-TeV in dilepton final states,''
  Phys.\ Lett.\ B {\bf 626}, 55 (2005)
  [hep-ex/0505082].
  %%CITATION = HEP-EX 0505082;%%

\bibitem{stop_exptRef}
V.~M.~Abazov {\it et al.}  [D0 Collaboration], 
D0-Notes 5050-CONF, 5039-CONF and 4866-CONF;
%\cite{Affolder:1999cz}
%\bibitem{Affolder:1999cz}
  A.~A.~Affolder {\it et al.}  [CDF Collaboration],
  %``Search for scalar top quark production in p anti-p collisions at  s**(1/2)
  %= 1.8-TeV,''
  Phys.\ Rev.\ Lett.\  {\bf 84}, 5273 (2000)
  [hep-ex/9912018].
  %%CITATION = HEP-EX 9912018;%%

\bibitem{left_snu_lsp} See, for example, 
A.~de Gouv\^ea, A.~Friedland and H.~Murayama,
  %``Less minimal supersymmetric standard model,''
  Phys.\ Rev.\ D {\bf 59}, 095008 (1999)
  [hep-ph/9803481];
  %%CITATION = HEP-PH 9803481;%%
A.~Datta, M.~Guchait and N.~Parua,
  %``Squark gluino mass limits revisited for nonuniversal scalar masses,''
  Phys.\ Lett.\ B {\bf 395}, 54 (1997)
  [hep-ph/9609413].
  %%CITATION = HEP-PH 9609413;%%

\bibitem{SUSYestab}
See, for example,   W.~Kilian, D.~Rainwater and J.~Reuter, 
  %``Distinguishing little-Higgs product and simple group models at the LHC and
  %ILC,''
  hep-ph/0609119;
  %%CITATION = HEP-PH 0609119;%%
A.~Alves, O.~Eboli and T.~Plehn,
  %``It's a gluino,''
  hep-ph/0605067;
  %%CITATION = HEP-PH 0605067;%%
A.~Datta, K.~Kong and K.~T.~Matchev,
  %``Discrimination of supersymmetry and universal extra dimensions at  hadron
  %colliders,''
  Phys.\ Rev.\ D {\bf 72}, 096006 (2005)
  [Erratum-ibid.\ D {\bf 72}, 119901 (2005)]
  [hep-ph/0509246];
  %%CITATION = HEP-PH 0509246;%%
and references therein. 

\end{thebibliography}
\end{document}